\documentclass[11pt,onecolumn,draftcls]{IEEEtran}
\usepackage{graphicx}
\usepackage{bm}
\usepackage{url}
\usepackage{amsfonts,amssymb,amsmath}
\usepackage{mystyle}
\usepackage{nomencl}
\allowdisplaybreaks                              
\makeatletter

\newcommand{\Rmnum}[1]{\expandafter\@slowromancap\romannumeral #1@}
\makeatother

\def\Pr{\mathbb P}
\def\Ex{\mathbb E} 
\def\bfm{\mathbf}
\def\sfm{\mathsf}
\def\calm{\mathcal}
\def\Nprimary {N}
\def\Mprimary {M}
\def\PSecondary {P}
\def\LowSINR {L}
\def\UpSINR {U}
\def\RmacInstant {R_{mac}}
\def\hmax {h_{max}}
\def\Smax {S_{max}}
\def\trace{\text{tr}}

\begin{document}

\title{Capacity Limits of Multiuser Multiantenna Cognitive
  Networks}
\date{\today} 
\author{{\large\em Yang Li and Aria
    Nosratinia}\\ The University of Texas at Dallas, Richardson, TX
  75080, USA \\ Email: liyang@student.utdallas.edu, aria@utdallas.edu}

\maketitle
\begin{abstract}
Unlike point-to-point cognitive radio, where the constraint
imposed by the primary rigidly curbs the secondary throughput,
multiple secondary users have the potential to more efficiently
harvest the spectrum and share it among themselves. This paper
analyzes the sum throughput of a multiuser cognitive radio system with
multi-antenna base stations, either in the uplink or downlink mode.
The primary and secondary have $\Nprimary $ and $n$ users,
respectively, and their base stations have $\Mprimary $ and $m$
antennas, respectively. We show that an {\em uplink secondary}
throughput grows with $\frac{m}{\Nprimary +1}\log n$ if the primary is
a downlink system, and grows with $\frac{m}{\Mprimary +1}\log n$ if
the primary is an uplink system. These growth rates are shown to be
optimal and can be obtained with a simple threshold-based user
selection rule. Furthermore, we show that the secondary throughput can
grow proportional to $\log n$ while simultaneously pushing the
interference on the primary down to zero, asymptotically. Furthermore,
we show that a {\em downlink secondary} throughput grows with $m\log
\log n$ in the presence of either an uplink or downlink primary
system. In addition, the interference on the primary can be made to go
to zero asymptotically while the secondary throughput increases
proportionally to $\log \log n$. Thus, unlike the point-to-point case,
multiuser cognitive radios can achieve non-trivial sum throughput
despite stringent primary interference constraints.
\end{abstract}

\section{Introduction}
\label{sec:introduction}

Currently, the spectrum assigned to licensed (primary) users is
heavily under-utilized~\cite{FCC}. Cognitive radio aims to improve the
utilization of spectrum by allowing cognitive (secondary) users to
access the same spectrum as primary users, as long as any performance
degradation of the primary users is tolerable.

In general, secondary users can access the spectrum via methods known
as overlay, interweave, and underlay~\cite{Jafar2009}. In the overlay
technique the secondary user not only transmits its own signal, but
also acts as a relay to compensate for its interference on the primary
user. The overlay method depends on the secondary transmitter having access to
primary's message~\cite{Devroye2006}.\footnote{Sometimes, this is
  referred to as an interference channel with degraded message sets.}
In the interweave technique~\cite{Mitola}, the secondary user first
senses spectrum holes and then transmits in the detected
holes. Reliable sensing in the presence of fading and shadowing has
proved to be challenging~\cite{Sahai2004}. Finally, in the underlay
technique~\cite{Ghasemi2007}, the secondary can transmit as long as
the interference caused on the primary is less than a pre-defined
threshold. The secondary user in this case is neither required to know
the primary user's message nor restricted to transmit in spectrum
holes.

This paper studies performance limits of an underlay cognitive network
consisting of multi-user and multi-antenna primary and secondary
systems. The primary and secondary systems are subject to mutual
interference, where the secondary has to comply with a set of
interference constraints imposed by the primary. We are interested in
the average sum rate (throughput) of the secondary system as the
number of secondary users grows. Moreover, we study how the secondary
throughput is affected by the size of primary network as well as the
severity of the interference constraints, which is one of the key
issues in the design of an underlay cognitive network.

A summary of the results of this paper is as follows. We assume that
the primary and secondary have $N$ and $n$ users, respectively, and
their base stations have $M$ and $m$ antennas, respectively.
\begin{itemize}
\item 
{\bfseries Secondary uplink (MAC):} the secondary average throughput
is shown to grow as $\Theta(\log n)$, which is achieved by a
threshold-based user selection rule. More precisely, the average
throughput of the secondary MAC channel grows as $\frac{m}{\Nprimary
  +1}\log n + O(1)$ when it coexists with the primary broadcast
channel, and grows as $\frac{m}{\Mprimary +1}\log n + O(1)$ when it
coexists with the primary MAC channel. By developing asymptotically
tight upper bounds, these growth rates are further proven to be
optimal. Moreover, the interference on the primary system can be
asymptotically forced to {\em zero}, while the secondary throughput
still grows as $\Theta(\log n)$. Specifically, for some non-negative
exponent $q$, the interference on the primary can be made to decline
as $\Theta(n^{-q})$, while the throughput of a secondary MAC grows as
$\frac{m-q\Nprimary}{\Nprimary +1}\log n + O(1)$ and
$\frac{m-q\Mprimary}{\Mprimary +1}\log n + O(1)$, respectively in
cases of primary broadcast and MAC channel.  The above results imply
that asymptotically the secondary system can attain a non-trivial
throughput {\em without} degrading the performance of the primary
system.

\item
{\bfseries Secondary downlink (broadcast):} the secondary average
throughput is shown to scale with $m\log \log n + O(1)$ in the
presence of either the primary broadcast or MAC channel. Hence, the
growth rate of throughput is unaffected (thus optimal) by the presence
of the primary system. In addition, the interference on the primary
can be asymptotically forced to {\em zero}, while maintaining the
secondary throughput as $\Theta(\log\log n)$. Specifically, for an
arbitrary exponent $0<q<1$, the interference can be made to decline as
$\Theta\big((\log n)^{-q}\big)$, while the secondary average
throughput grows as $m(1-q)\log \log n + O(1)$.

\end{itemize}

Some of the related earlier work is  as follows. Much of the past work
in the underlay cognitive radio involves point-to-point primary and
secondary systems. Ghasemi et al~\cite{Ghasemi2007} studies the
ergodic capacity of a point-to-point secondary link under various
fading channels. Multiple antennas at the secondary transmitter are
exploited by~\cite{Zhang2008} to manage the tradeoff between the
secondary throughput and the interference on the primary. In the
context of multi-user cognitive radios, Zhang et al~\cite{Zhang2009}
studies the power allocation of a single-antenna secondary system
under various transmit power constraints as well as interference
constraints.  Gastpar~\cite{Gastpar2007} studies the secondary
capacity via translating a receive power constraint into a transmit
power constraint.

Recently, ideas from opportunistic communication~\cite{Viswanath2002}
were used in underlay cognitive radios by selectively activating one
or more secondary users to maximize the secondary throughput while
satisfying interference constraints. The user selection in cognitive
radio is complicated because the secondary system must be mindful of
two criteria: the interference on the primary and the rate provided to
the secondary. Karama et al~\cite{Hamdi2009} selects secondary users
with channels almost orthogonal to a single primary user, so that the
interference on the primary is reduced. Jamal et al~\cite{Mitran2009,
  Mitran2010} obtains interesting scaling results for the sum rate by selecting
users causing the least interference. Some distinctions of our work
and~\cite{Mitran2009, Mitran2010} are worth noting.
First, Jamal et al~\cite{Mitran2009,Mitran2010} studies the hardening
of sum rate via convergence in probability, while we analyze the
average throughput, which requires a very different approach.%
\footnote{In general, convergence in probability does not imply
  convergence in any moment (thus average
  throughput)~\cite{Serflingbook}. For example, consider a sequence of
  rates $R_n=\log (1+X_n)$, where
\begin{equation}
X_n =\begin{cases}
     1 &\mbox{with probability $1-\frac{1}{n}$}\\ 
     \exp(n^2) &\mbox{with probability $\frac{1}{n}$ }
     \end{cases} \nonumber
\end{equation} 
Then, $\lim_{n\uparrow\infty}R_n=\log 2$ in probability, however,
$\lim_{n\uparrow\infty}\Ex[R_n]=\infty$ in probability. Therefore, the
average rate $\Ex[R_n]$ cannot be predicted based on the hardening (in
probability) of $R_n$.}
Second, we study a multi-antenna cognitive network
whereas~\cite{Mitran2009, Mitran2010} considers a single antenna
network. Third, we study the effect of the primary network size
(number of constraints) on the secondary throughput,
while~\cite{Mitran2009, Mitran2010} considers a single primary
constraint.

We use the following notation: $[\,\cdot\,]_{i,j}$ refers to the
$(i,j)$ element in a matrix, $|\cdot|$ refers to the cardinality of a
set or the Euclidean norm of a vector, $\diag(\cdot)$ refers to a
diagonal matrix, $\trace(\cdot)$ refers to the trace of a matrix, and
$I_{k\times k}$ refers to the $k\times k$ identity matrix. All $\log
(\cdot)$ is natural base. For any $\epsilon>0$, some positive $c_1$
and $c_2$, and sufficiently large $n$:
\begin{samepage}
\begin{align*}
&f(n)=O\big(g(n)\big):& |f(n)|&<c_1\,|g(n)|
  \\ &f(n)=\Theta\big(g(n)\big):& c_2\,|g(n)|<|f(n)|&<c_1\,|g(n)|
  \\ &f(n)=o\big(g(n)\big): & |f(n)|&<\epsilon\,|g(n)|
\end{align*}
\end{samepage}

We let $\calm{R}_{mac,w/o}^{opt}$ and $\calm{R}_{bc,w/o}^{opt}$ be the
{\em maximum} average throughput achieved by the secondary MAC and
broadcast channel {\em in the absence of} the primary,
respectively. In this case, we have regular MAC and broadcast
channels, and it is well known that $\calm{R}_{mac,w/o}^{opt}$ scales
as $m\log n$, and $\calm{R}_{bc,w/o}^{opt}$ scales as $m\log \log n$.

The remainder of this paper is organized as
follows. Section~\ref{sec:Model} describes the system model. The
average throughput of the secondary MAC channel is studied in
Section~\ref{sec:MAC}, where in Section~\ref{sec:Optimality} we prove
the achieved throughout is asymptotically optimal. The average
throughput of the secondary broadcast channel is investigated in
Section~\ref{sec:broadcast}. Numerical results are shown in
Section~\ref{sec:simulation}. Finally, Section~\ref{sec:conclusion}
concludes this paper.

\section{System Model}
\label{sec:Model}

\begin{figure}
\centering
\includegraphics[width=5in]{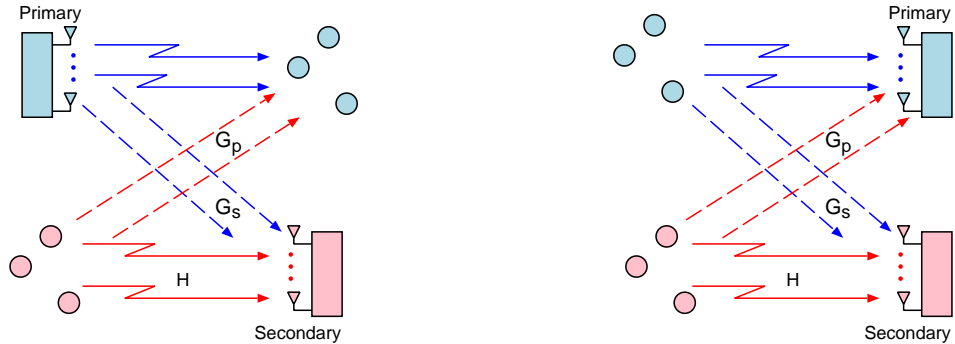}
\caption{Coexistence of the secondary MAC channel and the primary system}
\label{fig:ModelMAC}
\end{figure}

\begin{figure}
\centering
\includegraphics[width=5in]{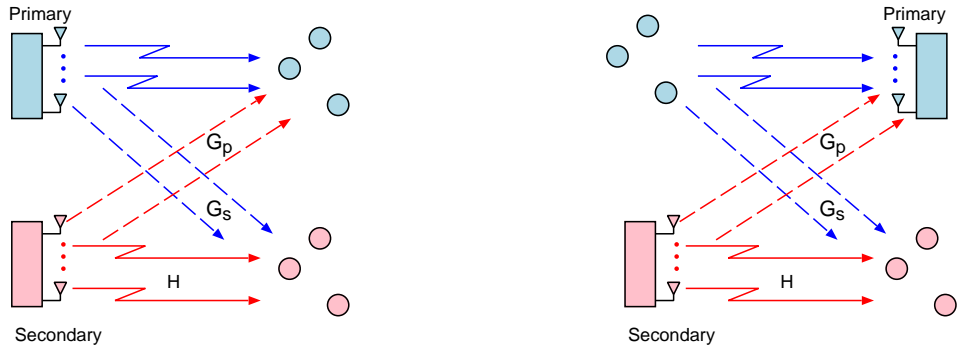}
\caption{Coexistence of the secondary broadcast channel and the primary system}
\label{fig:ModelBC}
\end{figure}

We consider a cognitive network consisting of a primary and a
secondary, each being either a MAC or broadcast channel
(Figure~\ref{fig:ModelMAC} and Figure~\ref{fig:ModelBC}). The primary
system has one base station with $\Mprimary $ antennas and $\Nprimary
$ users, while the secondary system consists of one base station with
$m$ antennas and $n$ users. The primary and secondary are subject to
mutual interference, which is treated as noise.  The secondary system
must comply with a set of interference power constraints imposed by
the primary. For simplicity of exposition, at the beginning primary
and secondary users (except base stations) are assumed to have one
antenna, however, as shown in the sequel, most of the results can be
directly extended to a scenario where each user has multiple antennas.

A block-fading channel model is assumed. All channel coefficients are
fixed throughout each transmission block, and are independent,
identically distributed (i.i.d.)
circularly-symmetric-complex-Gaussian with zero mean and unit
variance, denoted by $\calm{CN}(0,1)$. The secondary base station acts
as a scheduler: For each transmission block, a subset of the secondary
users is selected to transmit to (or receive from) the secondary base
station. We denote the collection of selected (active) secondary users
as $\calm{S}$. 

We begin by introducing a system model that applies to all four
scenarios in Figures~\ref{fig:ModelMAC} and~\ref{fig:ModelBC}, thus
simplifying notation in the remainder of the paper. The secondary
received signal is given by:
\begin{equation}
\bfm{y} = \bfm{H}(\calm{S})\,\bfm{x}_s + \bfm{G}_s\,\bfm{x}_p + \bfm{w} \label{eq:SUSignal}
\end{equation}
where $\bfm{y}$ represents the received signal vector, either signals
at a multi-antenna base station (uplink) or at different users
(downlink).  $\bfm{H}(\calm{S})$ is the channel coefficient matrix
between the active secondary users and their base station.
$\bfm{G}_s$ represents the cross channel coefficient matrix from the
primary transmitter(s) to the secondary receiver(s). The primary and
secondary transmit signal vectors are $\bfm{x}_p$ and $\bfm{x}_x$. The
variable $\bfm{w}$ is the received noise vector, where each entry of
$\bfm{w}$ is i.i.d. $\calm{CN}(0,1)$.

We assume both primary and secondary systems use Gaussian signaling,
subject to short-term power constraints. The transmit covariance
matrices of the primary and secondary systems are
\begin{equation}
Q_p = \Ex\big[\bfm{x}_p\bfm{x}^{\dag}_p\big]
\end{equation}
and
\begin{equation}
Q_s = \Ex\big[\bfm{x}_s\bfm{x}^{\dag}_s\big]
\end{equation}

When the secondary is a MAC channel, each secondary user is subject to
an individual short term power constraint $\rho_s$. The users do not
cooperate, therefore $Q_s$ is diagonal:
\begin{align}
Q_s =
\diag\big(\rho_1,\cdots,\rho_{|\calm{S}|}\big) \label{eq:SUTxSigM}
\end{align}
where $\rho_{\ell}\le \rho_s$, for $\ell=1,\cdots, |\calm{S}|$. In
this case, $\bfm{H}(\calm{S})$ has dimension $m\times
|\calm{S}|$. 

When the secondary is a broadcast channel, we assume the secondary base
station is subject to a short term power constraint $P_s$:
\begin{equation}
\trace(Q_s) \le P_s  \label{eq:SUTxSigB}
\end{equation}
In this case, $\bfm{H}(\calm{S})$ has dimension $|\calm{S}|\times
m$.

When the primary is a MAC channel, each primary user transmits
with power $\rho_p$ without user cooperation:
\begin{equation}
 Q_p = \rho_p \, I_{\Nprimary \times \Nprimary }  \label{eq:PUTxPowerM}
\end{equation}
Furthermore, each receive antenna at the primary base station can
tolerate interference with power $\Gamma$ from the secondary
system,\footnote{If each primary antenna or user tolerates a different
  interference power, the results of this paper still hold, as seen
  later.} that is
\begin{equation}
\big[\bfm{G}_p\, Q_s\, \bfm{G}_p^{\dag}\big]_{\ell,\ell} \le
\Gamma \label{eq:ConstraintsMM}
\end{equation}
for $\ell = 1,\cdots,\Mprimary $, where $\bfm{G}_p$ represents the cross
channel coefficient matrix from the secondary base station (or active
users) to the primary base station.

When the primary is a broadcast channel, the power constraint at the
primary base station is $\trace(Q_p)\le P_p$. For simplicity, we
assume\footnote{The asymptotic results remain the same, even if we
  allow $Q_p$ to be an arbitrary covariance matrix.}
\begin{equation}
 Q_p = \frac{P_p}{\Mprimary } \, I_{\Mprimary \times \Mprimary }  \label{eq:PUTxPowerB}
\end{equation}
Furthermore, each primary user tolerates interference with power $\Gamma$:
\begin{equation}
\big[\bfm{G}_p\, Q_s\, \bfm{G}_p^{\dag}\big]_{\ell,\ell} \le
\Gamma \label{eq:ConstraintsMB}
\end{equation}
for $\ell = 1,\cdots,\Nprimary $, where $\bfm{G}_p$ is the cross
channel coefficient matrix from the secondary base station (or active
users) to the primary users.

\section{Cognitive MAC Channel}
\label{sec:MAC}

Consider a MAC secondary in the presence of either a broadcast or MAC
primary. We wish to find how much throughput is available to the
secondary subject to rigid constraints on the secondary-on-primary
interference. We first construct a transmission strategy and find the
corresponding (achievable) average throughput. Then, we develop upper
bounds that are tight with respect to the throughput achieved.

The framework for the transmission strategy is as follows: For each
transmission block, the secondary base station determines an active
user set $\calm{S}$ as well as transmit power for all active users
$Q_s$.
For each transmission, from \eqref{eq:SUSignal}, the sum rate
(throughput) of the secondary system is:
\begin{align}
\RmacInstant & = \log \det \bigg(I + \bfm{H}(\calm{S}) Q_s
\bfm{H}^{\dag}(\calm{S}) + \bfm{G}_s Q_p
\bfm{G}_s^{\dag}\bigg) - \log \det \bigg(I + \bfm{G}_s
Q_p\bfm{G}_s^{\dag}\bigg) \label{eq:InstantR}
\end{align} 
subject to the interference constraints~\eqref{eq:ConstraintsMB}
and~\eqref{eq:ConstraintsMM} for the primary broadcast and MAC channel
respectively. 

The secondary average throughput is given by
\begin{equation}
\calm{R}_{mac} = \Ex [\RmacInstant ]
\end{equation}
For the development of upper bounds, we assume the secondary base
station knows all the channels. This is a genie-like argument that is
used solely for development of upper bounds. For the achievable
scheme, the requirement is more modest and is outlined after the
description of the achievable scheme (see Remark~\ref{remark:CSIM}).
\subsection{Achievable Scheme}

The objective is to choose $\calm{S}$ and $Q_s$, i.e., the secondary
active transmitters and their power, such that secondary throughput is
maximized subject to interference constraints on the primary.

The choice of $\mathcal{S}$ and $Q_s$ is coupled through the
interference constraints: either more secondary users can transmit
with smaller power, or fewer of them with higher power. We focus on a
simple power policy: All active secondary users transmit with the
maximum allowed power $\rho_s$. Hence, given an active user set
$\calm{S}$, we have
\begin{equation}
Q_s=\rho_sI_{|\calm{S}|\times |\calm{S}|} \label{eq:Q_sM}
\end{equation}
It will be shown that the on-off transmission (without any further
power adaptation) suffices to (asymptotically) achieve the maximum
average throughput. Furthermore, its simplicity facilitates analysis.

Recall that each primary user can tolerate interference with power
$\Gamma$. The interference on a primary user is guaranteed to be below
this level if $k_s$ secondary users are active, each causing
interference no more than $\alpha = \frac{\Gamma}{k_s}$. This bound allows
us to honor the interference constraints on the primary while
decoupling the action of different secondary users. Based on this
observation, we construct a user selection rule as follows.
First, we define an eligible secondary user set that disqualifies users
that cause too much interference on the primary:
\begin{equation}
\calm{A} = \begin{cases} \big\{i:
  \rho_s\big|[\bfm{G}_p]_{ji}\big|^2<\alpha,\ \text{for}\ j=1,\cdots, \Nprimary 
  \big\} & \mbox{\text{primary broadcast}}   
\\ \big\{i:\rho_s\big|[\bfm{G}_p]_{ji}\big|^2<\alpha,\ \text{for}\ j=1,\cdots, \Mprimary 
  \big\} & \mbox{\text{primary MAC}}   
\end{cases} \label{eq:EligibleSet}
\end{equation}
where $[\bfm{G}_p]_{ji}$ is the channel coefficient from the secondary
user $i$ to the primary user (antenna) $j$, and $\alpha$ is a pre-designed
interference quota. A secondary user is eligible if its interference
on each primary user (antenna) is less than $\alpha$.
Now, to satisfy the interference bound, we limit the number of secondary
transmitters to no more than $k_s$, where
\begin{equation}
k_s = \frac{\Gamma}{\alpha} \label{eq:k_s} 
\end{equation}
If $|\calm{A}|\le k_s$, then all eligible users can transmit. If
$|\calm{A}|> k_s$, then $k_s$ users will be chosen {\em randomly} from
among the eligible users to transmit.\footnote{Naturally the number of
  active users must be an integer, i.e., $\lfloor k_s\rfloor$. We do
  not carry the floor operation in the following developments for
  simplicity, noting that due to the asymptotic nature of the
  analysis, the floor operation has no effect on the final results.}
The number of eligible users, $|\calm{A}|$, is a random variable; the
number of active users is
\begin{equation}
|\calm{S}| = \min\big(k_s,|\calm{A}|\big) \label{eq:|S|}
\end{equation}
The transmission of $|\calm{S}|$ eligible users induces interference
no more than $\Gamma$ on any primary user or antenna. Notice that the
manner of user selection guarantees that the channel coefficients in
$\bfm{H}(\calm{S})$ remain independent and distributed as
$\calm{CN}(0,1)$.

Now we want to design an interference quota $\alpha$ to maximize the
secondary average throughput. Neither very small nor very large values
of $\alpha$ are useful within our framework: If $\alpha$ is very small, for
most transmissions few (if any) secondary users will be eligible, thus
the secondary throughput will be small. If $\alpha>\Gamma$, any transmitting
user might violate the interference constraint, so the secondary must
shut down (equivalently, we have $k_s<1$). The value of individual
interference constraint $\alpha$, or equivalently $k_s$, must be set
somewhere between these extremes.

Clearly, a desirable outcome would be to allow exactly the number of
users that are indeed eligible for transmission, i.e., $k_s\approx
|\calm{A}|$. But one cannot guarantee this in advance, because
$|\calm{A}|$ is a random variable. Motivated by this general insight,
we choose $\alpha$ such that
\begin{equation}
k_s=\Ex[|\calm{A}|] \label{eq:k_sDesigned}
\end{equation}
In Section~\ref{sec:Optimality}, we will verify that this choice of
$\alpha$ is enough to asymptotically achieve the maximum throughput.

\begin{remark} 
\label{remark:CSIM}
The above scheme does not require the secondary users to have full
channel knowledge. Each secondary user can compare its own cross
channel gains with a pre-defined interference quota $\alpha$, and then
decide its eligibility. After this, each eligible user can inform the
secondary base station via $1$-bit, so that the secondary base station
can determine $\calm{A}$ without knowing the cross channels from the
secondary users to the primary system. The secondary channels
$\bfm{H}(\calm{S})$ and the cross channels $\bfm{G}_s$ can be
estimated at the secondary base station. Therefore, this scheme can be
implemented with little exchange of channel knowledge.
\end{remark}

\subsection{Throughput Calculation}

\subsubsection{Secondary MAC with Primary Broadcast}

The primary base station transmits to $\Nprimary $ primary users,
where each user tolerates interference with power $\Gamma$. 
Notice that in~\eqref{eq:EligibleSet}, $[\bfm{G}_p]_{ji}$ is the
channel coefficient from the secondary user $i$ to the primary user
$j$ which is i.i.d. $\calm{CN}(0,1)$. Thus,
$\big|[\bfm{G}_p]_{ji}\big|^2$ is i.i.d. exponential. Therefore,
$|\calm{A}|$ is binomially distributed with parameter $(n,p)$, where
\begin{equation}
p = \big(1-e^{-\frac{\alpha}{\rho_s}}\big)^{\Nprimary } \label{eq:Prob0MB}
\end{equation}
For small $\frac{\alpha}{\rho_s}$, we have
\begin{equation}
p\approx\bigg(\frac{\alpha}{\rho_s}\bigg)^{\Nprimary }\label{eq:ProbMB}
\end{equation}
From~\eqref{eq:k_sDesigned}, the interference quota $\alpha$ is chosen
such that
\begin{equation}
k_s =np\approx n \bigg(\frac{\alpha}{\rho_s}\bigg)^{\Nprimary } \label{eq:k_s0MB}
\end{equation}
Substitute $\alpha = \frac{\Gamma}{k_s}$ into the above equation, and denote
the associated solution for $k_s$ as $\bar{k}_s$:
\begin{equation}
\bar{k}_s = \bigg(\frac{\Gamma}{\rho_s}\bigg)^{\frac{\Nprimary }{\Nprimary +1}} (n)^{\frac{1}{\Nprimary +1}} \label{eq:k_sMB}
\end{equation}
Thus, we can see $\Theta(n^{\frac{1}{\Nprimary +1}})$ secondary users
are allowed to transmit, and the interference quota is on the order of
$\Theta(n^{-\frac{1}{\Nprimary +1}})$. With the above choice of
interference quota, or the number of allowable active users, we state
one of the main results of this paper as follows.

\begin{theorem}
\label{thm:AvgRMB}
Consider a secondary MAC with a $m$-antenna base station and $n$ users
each with power constraint $\rho_s$. The secondary MAC operates in the
presence of a primary broadcast channel transmitting with power $P_p$
to $N$ users each with interference tolerance $\Gamma$. The secondary
average throughput satisfies:

\begin{align}
 \calm{R}_{mac} & \ge \frac{m}{\Nprimary +1}\log n + \frac{1}{\Nprimary +1}\log
 \big(\rho_s\Gamma^{\Nprimary }\big)- m\log(1+P_p) +
 O\big(n^{-\frac{1}{\Nprimary +1}}\log n \big) 
\\ \calm{R}_{mac}  & \le  \frac{m}{\Nprimary +1}\log n + \frac{1}{\Nprimary +1}\log \big(\rho_s\Gamma^{\Nprimary }\big)- \calm{R}_I + O\big(n^{-\frac{1}{\Nprimary +1}}\big)
\end{align}
with
\begin{equation}
\calm{R}_I = m_{\sfm{min}}\log \bigg(1 +
\frac{P_p}{\Mprimary }\exp\bigg(\frac{1}{m_{\sfm{min}}}\sum_{j=1}^{m_{\sfm{min}}}\sum_{i=1}^{m_{\sfm{max}}-j}\frac{1}{i}-\gamma\bigg)\bigg)
\end{equation}
where $m_{\sfm{min}} = \min(m,\Mprimary )$ and $m_{\sfm{max}} =
\max(m,\Mprimary )$. This throughput is achieved under the
threshold-based user selection with the choice of $\bar{k}_s$ given
by~\eqref{eq:k_sMB}.
\end{theorem}
\begin{Proof}
See Appendix~\ref{Appendix:thmMB}.
\end{Proof}

\begin{remark}
The essence of the above result is that the secondary average
throughput grows as $\frac{m}{\Nprimary +1}\log n + O(1)$, i.e.,
inversely proportional to the number of primary users. A noteworthy
special case is when the primary base station chooses to transmit to a
number of users equal to the number of its transmit antennas
($\Nprimary =\Mprimary $), a strategy which is known to be
near-optimum in terms of sum-rate~\cite{Caire2003}. Under this
condition:
\begin{equation}
\calm{R}_{mac} = \frac{m}{\Mprimary +1}\log n + O(1)  \nonumber
\end{equation}
Therefore, we have
\begin{equation}
\lim_{n\rightarrow\infty}\frac{\calm{R}_{mac}}{\calm{R}_{mac,w/o}^{opt}} = \frac{1}{\Mprimary +1} 
\end{equation}
where $\calm{R}_{mac,w/o}^{opt}$ is the maximum average throughput of
the secondary MAC {\em in the absence of} the primary
system. This ratio shows that the {\em compliance penalty} of the secondary
MAC system and its relationship with the characteristics of the
primary network.  
\end{remark}

\begin{remark}
The results in Theorem~\ref{thm:AvgRMB} can be directly extended to a
scenario where each primary user tolerates a different level of
interference. As long as all primary users allow non-zero interference
(no matter how small), we can let $\Gamma$ be the minimum allowable
interference, and the theorem still holds.
\end{remark}

So far we have analyzed the effect of small but constant primary
interference constraints, and shown that the secondary throughput
improves with increasing the number of secondary users. However, the
flexibility provided by the increasing number of secondary users can
be exploited not only to increase secondary throughput, but also to
reduce the primary interference. In fact, it is possible to
simultaneously suppress the interference on the primary down to {\em
  zero} while increasing the secondary throughput proportional to
$\log n$. The following corollary makes this idea precise:

\begin{corollary}
\label{cor:ReduceInterferenceMB}
Assuming the interference on each primary user is bounded as
$\Theta(n^{-q})$, the average secondary throughput satisfies
\begin{equation}
\calm{R}_{mac} = \frac{m-q\Nprimary }{\Nprimary +1} \log n
+ O(1)
\end{equation}
where $0< q < \frac{m}{\Nprimary }$.
\end{corollary}
\begin{Proof}
Because the proof of Theorem~\ref{thm:AvgRMB} holds for $\Gamma =
\Theta(n^{-q})$, the corollary follows by substituting
$\Gamma=\Theta(n^{-q})$ into the lower and upper bounds given by
Theorem~\ref{thm:AvgRMB}.
\end{Proof}

\begin{remark}
\label{remark:qMB}
The corollary above explores a tradeoff where primary interference is
made to decrease polynomially, i.e., proportional to $n^{-q}$. We saw
that this leads to a secondary sum rate that decreases linearly in
$q$. If we reduce the primary interference more slowly, i.e.,
decreasing as $\Theta(\frac{1}{\log n})$, the growth rate of secondary
sum-rate will behave as though the primary interference constraint is
fixed. Conversely, if we try to suppress the primary interference
faster than $\Theta(n^{-q})$, the secondary throughput will
asymptotically remain stagnant or will go to zero.

\end{remark}


\subsubsection{Secondary MAC with a Primary MAC}

Recall that each antenna at the primary base station allows
interference with power $\Gamma$. By regarding each antenna of the primary
base station as a virtual user, we can re-use most of the analysis
that was developed in the previous section. Thus, the steps leading to
Eq.~\eqref{eq:k_sMB} can be repeated to obtain the number of allowable
active secondary users:
\begin{equation}
\bar{k}_s = \bigg(\frac{\Gamma}{\rho_s}\bigg)^{\frac{\Mprimary }{\Mprimary +1}}
(n)^{\frac{1}{\Mprimary +1}} \label{eq:k_sMM}
\end{equation}
With this allowable active users $\bar{k}_s$ and slight modifications,
we obtain a result that parallels Theorem~\ref{thm:AvgRMB}.

\begin{theorem}
\label{thm:AvgRMM}
Consider a secondary MAC with a $m$-antenna base station and $n$ users
each with power constraint $\rho_s$. The secondary MAC operates in the
presence of a primary MAC channel where each user transmits with power
$\rho_p$ to a $\Mprimary$-antenna base station with interference
tolerance $\Gamma$ on each antenna. The secondary average throughput
satisfies:
\begin{align}
 \calm{R}_{mac} & \ge \frac{m}{\Mprimary +1}\log n + \frac{1}{\Mprimary +1}\log
 \big(\rho_s\Gamma^{\Mprimary }\big)- m\log(1+\rho_p\Nprimary ) +
 O\big(n^{-\frac{1}{\Mprimary +1}}\log n \big) 
\\ \calm{R}_{mac} & \le
 \frac{m}{\Mprimary +1}\log n + \frac{1}{\Mprimary +1}\log \big(\rho_s\Gamma^{\Mprimary }\big)-
 \calm{R}_I + O\big(n^{-\frac{1}{\Mprimary +1}}\big)
\end{align}
with
\begin{equation}
\calm{R}_I = m_{\sfm{min}}\log \bigg(1 +
\rho_p\exp\bigg(\frac{1}{m_{\sfm{min}}}\sum_{j=1}^{m_{\sfm{min}}}\sum_{i=1}^{m_{\sfm{max}}-j}\frac{1}{i}-\gamma\bigg)\bigg)
\end{equation}
where $m_{\sfm{min}} = \min(m,\Nprimary )$ and $m_{\sfm{max}} =
\max(m,\Nprimary )$. This throughput is achieved under the
threshold-based user selection with the choice of $\bar{k}_s$ given
by~\eqref{eq:k_sMM}.
\end{theorem}

A tradeoff exists between the primary interference reduction
and the secondary throughput enhancement, which is stated by the
following corollary. All the remarks made after
Corollary~\ref{cor:ReduceInterferenceMB} are applicable here.

\begin{corollary}
\label{cor:ReduceInterferenceMM}
Assuming the interference on each antenna of the primary base station
is bounded as $\Theta(n^{-q})$, the average secondary throughput
satisfies
\begin{equation}
\calm{R}_{mac} = \frac{m-q\Mprimary }{\Mprimary +1} \log n
+ O(1)
\end{equation}
where $0< q < \frac{m}{\Mprimary }$.
\end{corollary}

\subsection{Upper Bounds for Secondary Throughput}
\label{sec:Optimality}

So far we have seen achievable rates of a cognitive MAC channel in the
presence of either a primary broadcast or MAC. We now develop
corresponding upper bounds.

\begin{theorem}
\label{thm:Optimality}
Consider a secondary MAC with a $m$-antenna base station and $n$
users. The {\em maximum} average throughput of the secondary,
$\calm{R}_{mac}^{opt}$, satisfies
\begin{equation}
 \calm{R}_{mac}^{opt} \le \frac{m}{\Nprimary +1} \log n + O(\log\log n)
\end{equation}
in the presence of a primary broadcast channel transmitting to
$\Nprimary$ users. Similarly, $\calm{R}_{mac}^{opt}$ satisfies
\begin{equation}
\calm{R}_{mac}^{opt} \le \frac{m}{\Mprimary +1} \log n + O(\log\log n) 
\end{equation}
in the presence of a primary MAC, where each user transmits to a
$\Mprimary$-antenna base station.
\end{theorem}
\begin{Proof}
See Appendix~\ref{Appendix:thmOpt}.
\end{Proof}

\begin{remark}
By comparing the upper bounds with the achievable rates obtained by
the thresholding strategy, we see that the achievable rates are at
most $O(\log \log n)$ away from the upper bounds, a difference which
is negligible relative to the dominant term $\Theta(\log n)$. Thus,
the growth of the {\em maximum} average throughput of a cognitive MAC
is $\frac{m}{\Nprimary +1} \log n$ in the presence of the primary
broadcast channel, and $\frac{m}{\Mprimary +1} \log n$ in the presence
of the primary MAC channel. Both the achievable rates and the upper
bounds show that the average cognitive sum-rate is inversely
proportional to the number of primary-imposed constraints,
asymptotically.
\end{remark}

\subsection{Discussion}

Recall that our method determines eligible cognitive MAC users based
on their cross channel gains. To satisfy the interference constraints,
our selection rule then allows $\Theta(n^{\frac{1}{\Nprimary +1}})$,
or $\Theta(n^{\frac{1}{\Mprimary +1}})$, of these users to be active
simultaneously, in the presence of either the primary broadcast or
MAC. If there are more eligible users than the allowed number, we
choose from among the eligible users randomly. In this process, the
forward channel gain of the cognitive users does not come into play,
and still an optimal growth rate is achieved. This can be intuitively
explained as follows. The total received signal power at the cognitive
base station grows linearly with the number of active users, and the
total received signal power determines the sum rate. On the other
hand, selecting good cognitive users according to their secondary
channel strengths can only offer logarithmic power gains (with respect
to $n$)~\cite{Viswanath2002}, which is negligible compared to the
linear gains due to increasing the number of active users. Therefore
the cross channel gains are more important in this case.\footnote{In a
  somewhat different context, the work of Jamal et
  al.~\cite{Mitran2010} also indicates that cross channels can be more
  important than the forward channels.}  Note that we do not imply
that knowledge of the cognitive forward channel is useless; our
conclusion only says that once the cross channels are taken into
account, the {\em asymptotic growth} of the secondary throughput
cannot be improved by any use of the cognitive forward channel.

Although we have allowed the base stations to have multiple antennas,
so far the users have been assumed to have only one antenna. We now
consider a generalization to the case where all users have multiple
antennas. Consider a secondary MAC in the presence of a primary
broadcast, where each primary and secondary user have $t_p$ and $t_s$
antennas respectively. We apply a separate interference constraint on
each antenna of each primary user, which guarantees the satisfaction
of the overall interference constraint on any primary user. On each of
the $t_s$-antenna secondary users, we shall allocate $t_s-1$ degrees
of freedom for zero-forcing and only one degree of freedom for
cognitive transmission. Using this strategy, we can ensure that
$t_s-1$ of the receive antennas on the primary are exempt from
interference. Thus, the total number of interference constraints will
reduce from $t_p \Nprimary $ to $ t_p\Nprimary+1-t_s$. By using an
analysis similar to the development of Theorem~\ref{thm:AvgRMB}, one
can show that the growth rate $\frac{m\log n}{\max(1,\,
  t_p\Nprimary+2-t_s)}$ is achievable. For the converse, the situation
is more complicated, because here the correlation among the antennas
of the secondary users must be accounted for. Nevertheless, in some
cases it is possible to show without much difficulty that the above
achieved throughput is indeed asymptotically optimal. For example, in
the presence of the primary MAC, if $t_s>\Mprimary $, the secondary
MAC channel can have a throughput that grows as $m\log n$ by letting
each active secondary user completely eliminate the interference on
the primary. Similarly, in the presence of a primary broadcast
channel, if $t_s>t_p\Nprimary$, the secondary MAC channel can also
have a throughput that grows as $m\log n$. The achieved growth rate is
optimal because it coincides with the the growth rate of
$\calm{R}_{mac,w/o}^{opt}$, which is always an upper bound.

\section{Cognitive Broadcast Channel}
\label{sec:broadcast}

\subsection{Achievable Scheme}

We consider a random beam-forming technique where the secondary base
station opportunistically transmits to $m$ secondary users
simultaneously~\cite{Sharif2005}. Specifically, the secondary base
station constructs $m$ orthonormal beams, denoted by
$\{\bfm{\phi}_j\}_{j=1}^{m}$, and assigns each beam to a secondary
user. Then, the secondary base station broadcasts to $m$ selected
users. The selection of users and beam assignment will be addressed
shortly.

Considering an equal power allocation among $m$ users, the transmitted
signal from the secondary base station is given by:
\begin{equation}
\bfm{x}_s = \sum_{j=1}^{m} \sqrt{\frac{\PSecondary}{m}}\;\bfm{\phi}_j\;
x_j \label{eq:SUSignalB}
\end{equation}
where $\bfm{\phi}_j$ is the beam-forming vector $j$ with dimension
$m\times 1$, $x_j$ is the signal transmitted along the beam $j$, and
$\PSecondary$ is the total transmit power. In this case, we have
\begin{equation}
Q_s = \frac{\PSecondary}{m}I_{m\times m} \label{eq:SUTxRandomB}
\end{equation}
Notice that $\PSecondary$ is subject to the power constraint $P_s$ as
well as a set of interference constraints imposed by the
primary. Thus, the value of $\PSecondary$ depends on the cross
channels from the secondary base station to the primary system.

Assuming the beam $j$ is assigned to user
$i$. From~\eqref{eq:SUSignal} and~\eqref{eq:SUSignalB}, the received
signal at the secondary user $i$ is given by
\begin{equation}
y_i = \bfm{h}^{\dag}_i\bfm{\phi}_j x_j +  \sum_{k\ne
  j}\bfm{h}^{\dag}_i \bfm{\phi}_kx_k + \bfm{g}_{s,i}^{\dag}\bfm{x}_p + w_i
\end{equation}
where $\bfm{h}^{\dag}_i$ is the $1\times m$ vector of channel
coefficient from the secondary base station to the secondary user $i$,
and $\bfm{g}_{s,i}^{\dag}$ is the $1\times \Mprimary $ (or
$1\times \Nprimary $) vector of channel coefficients from the primary
base station (or users) to the secondary user $i$. The received
signal-to-noise-plus-interference-ratio (SINR) at the secondary user
$i$ (with respect to  beam $j$) is
\begin{equation}
\sfm{SINR}_{i,j}= \frac{\frac{\PSecondary}{m}|\bfm{h}^{\dag}_i\bfm{\phi}_j|^2}{1+\frac{\PSecondary}{m}\sum_{k\ne
  j}|\bfm{h}^{\dag}_i \bfm{\phi}_k|^2 +
  \bfm{g}^{\dag}_{s,i}\,Q_p \,\bfm{g}_{s,i}} \label{eq:SINRB}
\end{equation}

The random beam technique assigns each beam to the secondary user that
results in the highest SINR. Because the probability of more
than two beams being assigned to the same secondary user is
negligible, we have~\cite{Sharif2005}
\begin{align}
\calm{R}_{bc} &\approx \Ex\bigg[\sum_{j=1}^m\log \big(1+\max_{1\le
    i\le n}\sfm{SINR}_{i,j}\big)\bigg]
\\ & =  m\Ex\bigg[\log \big(1+\max_{1\le
    i\le n}\sfm{SINR}_{i,j}\big)\bigg]
\label{eq:AvgRB}
\end{align}
The above analysis holds in the presence of either the primary
broadcast or MAC channel; the only difference is the constraints on
$\PSecondary$ and $Q_p$.  Since the SINR is symmetric across all
beams, the subscript $j$ will be omitted in the following analysis.

\begin{remark}
\label{remark:CSIB}
We briefly address the issue of channel state information. All users
are assumed to have receiver side channel state information.  On the
transmit side, the secondary base station does not need to have full
channel knowledge; only the SINR is needed. Each secondary user can
estimate its own SINR with respect to each beam, and feed it back to
the secondary base station~\cite{Sharif2005}. Based on collected SINR,
the secondary base station performs user selection. The secondary base
station needs to know $\bfm{G}_p$ to adjust $\PSecondary$ such that
the interference constraints on the primary are satisfied.
\end{remark}

\subsection{Throughput Calculation}

\subsubsection{Secondary Broadcast with Primary Broadcast} 
The secondary system has to comply with the constraints on $\Nprimary
$ primary users. To maximize the throughput, the secondary base
station transmits at the maximum allowable
power. From~\eqref{eq:ConstraintsMB} and~\eqref{eq:SUTxRandomB}, we
have
\begin{equation}
\PSecondary = \min \big( \frac{m\Gamma}{|\bfm{g}^{\dag}_{p,1}|^2},\cdots,
\frac{m\Gamma}{|\bfm{g}^{\dag}_{p,\Nprimary }|^2},P_s\big) \label{eq:SUTxPowerBB}
\end{equation}
where $\bfm{g}^{\dag}_{p,\ell}$ is the row $\ell$ of $\bfm{G}_p$.
Then, we substitute $Q_p$ given by~\eqref{eq:PUTxPowerB}
into~\eqref{eq:SINRB}, and obtain the SINR at the secondary user $i$
with respect to the beam $j$:
\begin{equation}
\sfm{SINR}_i = \frac{|\bfm{h}^{\dag}_i\bfm{\phi}_j|^2}{\frac{m}{\PSecondary}+\sum_{k\ne
  j}|\bfm{h}^{\dag}_i \bfm{\phi}_k|^2 +
  \frac{mP_p}{\Mprimary \PSecondary}|\bfm{g}_{s,i}|^2} \label{eq:SINRBB}
\end{equation}

Our analysis of $\max_i \sfm{SINR}_i$, which is required to evaluate
the throughput in Eq.~\eqref{eq:AvgRB}, does not
follow~\cite{Sharif2005} because the denominator involves a sum of two
Gamma distributions with different scale parameters: $\sum_{k\ne
  j}|\bfm{h}^{\dag}_i \bfm{\phi}_k|^2$ has Gamma$(m-1,1)$ and $
\frac{mP_p}{\Mprimary \PSecondary}|\bfm{g}_{s,i}|^2$ has
Gamma$(\Mprimary,\frac{mP_p}{\Mprimary \PSecondary})$. Fortunately,
lower and upper bounds can be leveraged to simplify the analysis. We
define:
\begin{equation}
\theta = \frac{mP_p}{\Mprimary  \PSecondary}
\end{equation}
We consider the case when $\frac{mP_p}{\Mprimary P_s}\ge 1$.  The
techniques can then be generalized to the case of
$\frac{mP_p}{\Mprimary P_s} < 1$.\footnote{When $\frac{mP_p}{\Mprimary
    P_s} < 1$, one can define $\theta =\max(\frac{mP_p}{\Mprimary
    P},1)$. Then, we can use Bayesian expansion via conditioning on
  $\{\PSecondary<\frac{mP_p}{\Mprimary }\}$ and its complement, where
  both conditional terms can be shown to have the same growth rate.}
When  $\frac{mP_p}{\Mprimary P_s}\ge 1$, we have $\theta \ge 1$ for all $\PSecondary$.
We define:
\begin{equation}
\LowSINR_i=
\frac{|\bfm{h}^{\dag}_i\bfm{\phi}_j|^2}{\frac{m}{\PSecondary}+\theta \big(\sum_{k\ne
    j}|\bfm{h}^{\dag}_i \bfm{\phi}_k|^2 + |\bfm{g}_{s,i}|^2\big)}
\end{equation}
and
\begin{equation}
\UpSINR_i=  \frac{|\bfm{h}^{\dag}_i\bfm{\phi}_j|^2}{\frac{m}{\PSecondary} + \theta |\bfm{g}_{s,i}|^2}
\end{equation}
where $\LowSINR_i$ and $\UpSINR_i$ are random variables that depend on
channel realizations. Conditioned on $P$, the denominators of
$\LowSINR_i$ and $\UpSINR_i$ have Gamma distributions, which
simplifies the analysis. 

For $1\le i \le n$, we have
\begin{equation}
\LowSINR_i \le \sfm{SINR}_i \le \UpSINR_i
\end{equation}
Hence,
\begin{equation}
\LowSINR_{max} \le \max_{1\le i \le n} \sfm{SINR}_i \le \UpSINR_{max}
\end{equation}
where $\LowSINR_{max} = \max_i\LowSINR_i$ and $\UpSINR_{max} = \max_i\UpSINR_i$.
Therefore for any $x$, we have
\begin{equation}
\Pr(\LowSINR_{max}>x)\le \Pr(\max_{1\le i \le n}\sfm{SINR}_i>x)\le \Pr(\UpSINR_{max}>x)
\end{equation}
which implies~\cite{Moshe} that $\max_i\sfm{SINR}_i$ is
stochastically greater than $\LowSINR_{max}$, but stochastically
smaller than $\UpSINR_{max}$. We now use the following fact about
stochastic ordering:
\begin{lemma}[\cite{Moshe}]
\label{lemma:StochasticOrder}
If random variable $X$ is stochastically smaller than $Y$ and
$h(\cdot)$ is an increasing function, assuming $h(X)$ and $h(Y)$ are
measurable according to their distributions:
\begin{equation}
\Ex[h(X)] \le \Ex[h(Y)]
\end{equation}
\end{lemma}

Based on the above lemma, the secondary average throughput is bounded
as follows:
\begin{equation}
m\Ex\big[ \log (1+\LowSINR_{max})\big] \le
\calm{R}_{bc}\le m \Ex\big[\log (1+\UpSINR_{max})\big] \label{eq:AvgRBounds}
\end{equation}
We study the lower and upper bounds given by~\eqref{eq:AvgRBounds},
instead of directly analyzing $\calm{R}_{bc}$. Some useful properties
of $\LowSINR_{max}$ and $\UpSINR_{max}$ are as follows.

\begin{lemma}
\label{lemma:SINRBoundsBB}
Conditioned on $\PSecondary=\rho$, 
\begin{align}
&\Pr\bigg(\LowSINR_{max} \ge b_n -\frac{\rho}{m}\log\log n \,\bigg |\,
  \PSecondary=\rho \bigg) = 1-
  \Theta\bigg(\frac{1}{n}\bigg) \label{eq:L1BB} 
\\& \Pr\bigg(\UpSINR_{max}  <
  d_n + \frac{\rho}{m}\log\log n \,\bigg |\,\PSecondary=\rho \bigg) = 1 - \Theta\bigg(\frac{1}{\log
    n}\bigg) \label{eq:U2BB}
\\ & \Ex\bigg[ \UpSINR_{max} \,\bigg|\, \UpSINR_{max} > d_n+\frac{\rho}{m}\log\log n
  ,\PSecondary=\rho\bigg] < O(n\log n) \label{eq:U2ConditionalBB}
\end{align}
where $b_n = \frac{\rho}{m}\log n - \frac{\rho(m+\Mprimary -1)}{m} \log\log n
+O\big(\log\log\log n\big) $ and $d_n = \frac{\rho}{m}\log n -
\frac{\rho \Mprimary }{m} \log\log n +O\big(\log\log\log n\big) $.
\end{lemma}
\begin{Proof}
See Appendix~\ref{Appendix:lemmaSINR}.
\end{Proof}

Based on the above two lemmas, we obtain the following results for the
secondary throughput:
\begin{theorem}
\label{thm:AvgRBB}
Consider a secondary broadcast channel with $n$ users and a
$m$-antenna base station with power constraint $P_s$. The secondary
broadcast operates in the presence of a primary broadcast channel
transmitting with power $P_p$ to $\Nprimary$ users each with
interference tolerance $\Gamma$. The secondary average throughput
satisfies:
\begin{align}
 \calm{R}_{bc} &> m \log\big(\Gamma \log   n \big) -m\log\big(\tilde{\mu}_1 +
\frac{m\Gamma}{P_s}\big) + O\big(\frac{\log\log n}{\log n}\big) \nonumber
\\ \calm{R}_{bc} &< m\log(\Gamma\log n) - m\log \tilde{\mu}_2 + O(1) \nonumber
\end{align}
where $\tilde{\mu}_1 = \Ex[\max_{1\le i \le \Nprimary } |\bfm{g}_{p,i}^{\dag}|^2 ]$ and
$\tilde{\mu}_2 =\big(\Ex\big[ 1/\max_{1\le i \le \Nprimary } |\bfm{g}_{p,i}^{\dag}|^2
  \big]\big)^{-1}$.
\end{theorem}
\begin{Proof}
See Appendix~\ref{Appendix:thmBB}.
\end{Proof}

\begin{remark}
\label{remark:remark1BB}
The result above states that $\calm{R}_{bc}=m\log \log n + O(1)$, thus
\begin{equation}
\lim_{n\rightarrow\infty} \frac{\calm{R}_{bc}}{\calm{R}_{bc,w/o}^{opt}} = 1 \label{eq:RbcLimit}
\end{equation}
where $\calm{R}_{bc,w/o}^{opt}$ is the maximum average throughput of
the secondary broadcast channel {\em in the absence of} the primary
system. Therefore, the achieved average throughput is {\em
  asymptotically optimal}, because we always have $\calm{R}_{bc}\le
\calm{R}_{bc,w/o}^{opt}$. Thus, we have a positive result: The growth
rate of the secondary average throughput is unaffected by the
constraints and interference imposed by the primary, as long as each
primary user tolerates some small but fixed interference.
\end{remark}

The above results naturally lead to the question: How small can we
make the interference on the primary, while still having a secondary
average throughput that grows as $\Theta(\log\log n)$. We find that
$\Gamma$, the interference on each primary user, can
asymptotically go to {\em zero}, as shown by the next corollary.

\begin{corollary}
\label{cor:ReduceInterferenceBB}
Assuming the interference on each primary user is bounded as
$\Theta\big((\log n)^{-q}\big)$, the average secondary throughput
satisfies:
\begin{equation}
\calm{R}_{bc} = (1-q)m\log\log n + O(1)
\end{equation}
where $0< q < 1$.
\end{corollary}

\begin{remark}
\label{remark:BB}
Reducing the interference on the order of $\Theta\big((\log
n)^{-q}\big)$ sheds lights on how fast the interference can be reduced
on the primary, while having a non-trivial secondary throughout. For
$q> 1$, it does not imply $\calm{R}_{bc}$ is zero or negative; it only
means that $\calm{R}_{bc}$ is on the order of $o(\log\log n)$. Slower
interference reduction, e.g.  proportional to $\Theta\big((\log \log
n)^{-1}\big)$, will give maximal asymptotic growth of secondary
throughput, i.e., $m\log\log n$.
\end{remark}


\subsubsection{Secondary Broadcast with Primary MAC}
The analysis of this case closely parallels the analysis of the
primary broadcast. The secondary transmit power is given by
\begin{equation}
\PSecondary = \min \big( \frac{m\Gamma}{|\bfm{g}^{\dag}_{p,1}|^2},\cdots,
\frac{m\Gamma}{|\bfm{g}^{\dag}_{p,\Mprimary }|^2},P_s\big) \label{eq:SUTxPowerBM}
\end{equation}
where $\bfm{g}^{\dag}_{p,\ell}$ is the row $\ell$ of $\bfm{G}_p$. The
MAC primary system produces power $\Nprimary \rho_p$ and has
$\Mprimary$ interference constraints. From the viewpoint of the
secondary, this is all the information that is needed. Therefore the
analysis of Theorem~\ref{thm:AvgRBB} can be essentially repeated to
obtain the following result.
\begin{theorem}
\label{thm:AvgRBM}
Consider a secondary broadcast channel with $n$ users and a
$m$-antenna base station with power constraint $P_s$. The secondary
broadcast operates in the presence of a primary MAC where each user
transmits with power $\rho_p$ to a $\Mprimary$-antenna base station
with interference tolerance $\Gamma$ on each antenna. The secondary
average throughput satisfies:
\begin{align}
 \calm{R}_{bc} &> m \log\big(\Gamma \log   n \big) -m\log\big(\tilde{\mu}_3 +
\frac{m\Gamma}{P_s}\big) + O\big(\frac{\log\log n}{\log n}\big) \nonumber
\\ \calm{R}_{bc} &< m\log(\Gamma\log n) - m\log \tilde{\mu}_4 + O(1) \nonumber
\end{align}
where $\tilde{\mu}_3 = \Ex[\max_{1\le i \le \Mprimary } |\bfm{g}_{p,i}^{\dag}|^2 ]$ and
$\tilde{\mu}_4 =\big(\Ex\big[ 1/\max_{1\le i \le \Mprimary } |\bfm{g}_{p,i}^{\dag}|^2
  \big]\big)^{-1}$.
\end{theorem}

\begin{remark}
Theorem~\ref{thm:AvgRBB} and Theorem~\ref{thm:AvgRBM} can be extended
to a scenario where each primary and secondary user has multiple
antennas. A straightforward way is to regard each primary and
secondary antenna as a virtual user. Using an analysis similar to the
single-antenna case, the secondary broadcast channel can be shown to
achieve a throughput scaling as $m\log \log n$ (thus optimal). The
details are straight forward and are therefore omitted for brevity.
\end{remark}

Similar to Corollary~\ref{cor:ReduceInterferenceBB}, we can also
obtain the tradeoff between the primary interference reduction and the
secondary throughput enhancement as follows. All the remarks following
Corollary~\ref{cor:ReduceInterferenceBB} apply to the present case as
well.
\begin{corollary}
\label{cor:ReduceInterferenceBM}
Assuming the interference on each antenna of the primary base station
is bounded as $\Theta\big((\log n)^{-q}\big)$, the average secondary
throughput satisfies:
\begin{equation}
\calm{R}_{bc} = (1-q)m\log\log n + O(1)
\end{equation}
where $0< q < 1$.
\end{corollary}

\section{Numerical Results}
\label{sec:simulation}

In this section, we concentrate on numerical results in the presence
of the primary broadcast channel; the results in the presence of the
primary MAC channel are similar thus omitted. For all simulations, we
consider: $P_p=P_s=\rho_s=5$, the secondary base station has $m=4$
antennas, and the primary base station has $\Mprimary =2$ antennas and
the number of primary users is $\Nprimary =2$.

\begin{figure}
\centering
\includegraphics[width=3.5in]{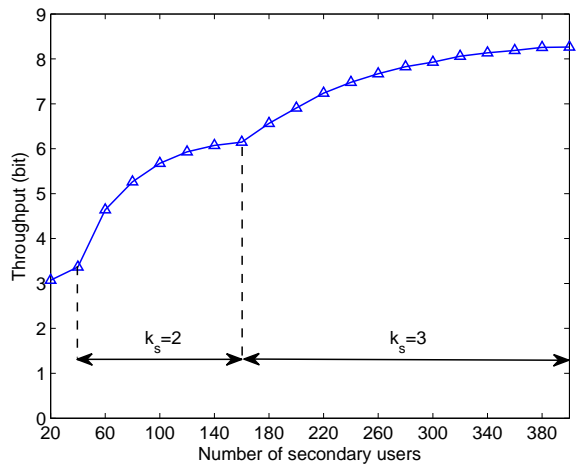}
\caption{Secondary MAC: Throughput versus user number ($\Gamma=2$)}
\label{fig:MB}
\end{figure}

\begin{figure}
\centering
\includegraphics[width=3.5in]{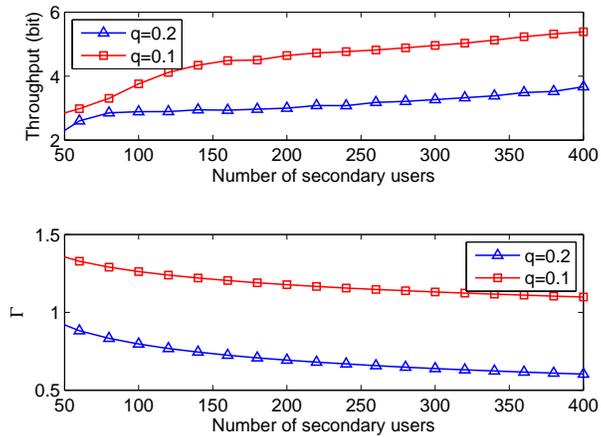}
\caption{Secondary MAC: Throughput versus user number ($\Gamma = 2n^{-q} $)}
\label{fig:MBInt}
\end{figure}

Figure~\ref{fig:MB} illustrates the secondary average throughput given
by Theorem~\ref{thm:AvgRMB}. The allowable interference power on each
primary user is $\Gamma=2$. The slope of the throughput curve is
discontinuous at some points, because the allowable number of active
secondary users must be an integer $\lfloor k_s\rfloor$ (also
see~Eq.\eqref{eq:k_s0MB}). As mentioned earlier, the floor operation
does not affect the asymptotic results.  Figure~\ref{fig:MBInt}
presents the tradeoff between the tightness of the primary constraints
and the secondary throughput, as shown by
Corollary~\ref{cor:ReduceInterferenceMB}. The interference power
constraint $\Gamma$ is $2n^{-q}$ for $q=0.1$ and $0.2$
respectively. 
As expected, for $q=0.2$ the interference on primary decreases faster
than $q=0.1$ and the secondary throughput increases more slowly.

\begin{figure}
\centering
\includegraphics[width=3.5in]{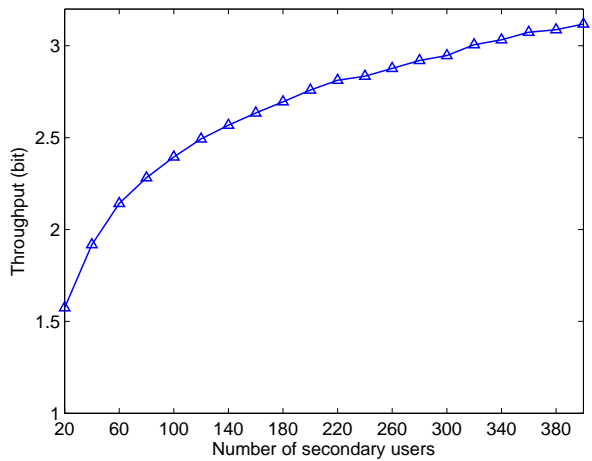}
\caption{Secondary broadcast: Throughput versus user number ($\Gamma=2$)}
\label{fig:BB}
\end{figure}

\begin{figure}
\centering
\includegraphics[width=3.5in]{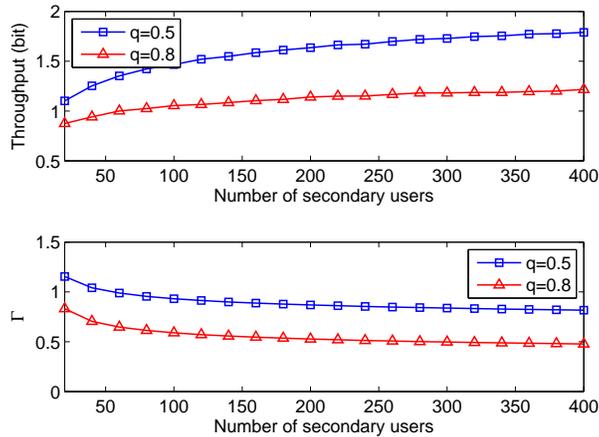}
\caption{Secondary broadcast: Throughput versus user number ( $\Gamma=2(\log n)^{-q}$)}
\label{fig:BBInt}
\end{figure}

Figure~\ref{fig:BB} shows the secondary throughput versus the number
of secondary users in the presence of the primary broadcast channel
(Theorem~\ref{thm:AvgRBB}), where the interference power is $\Gamma=2$. In
Figure~\ref{fig:BBInt}, we show the tradeoff between the secondary
throughput and the interference on the primary, as described in
Corollary~\ref{cor:ReduceInterferenceBB}. We set $\Gamma$ to decline as
$2(\log n)^{-q}$, for $q=0.5$ and $q=0.8$, respectively. Clearly, for
$q=0.5$, the interference power decreases faster than
$q=0.8$, while the secondary throughput increases more slowly.

\section{Conclusion}
\label{sec:conclusion}

In this paper, we study the performance limits of an underlay
cognitive network consisting of a multi-user and multi-antenna
primary and secondary systems. We find the average throughput limits
of the secondary system as well as the tradeoff between this
throughput and the tightness of constraints imposed by the primary
system. Given a set of interference power constraints on the primary,
the maximum average throughput of the secondary MAC grows as
$\frac{m}{\Nprimary +1}\log n$ (primary MAC), and $\frac{m}{\Mprimary +1}\log n$
(primary broadcast). These growth rates are attained by the simple
threshold-based user selection rule. Interestingly, the secondary
system can force its interference on the primary to zero while
maintaining a growth rate of $\Theta(\log n)$. For the secondary
broadcast channel, the secondary average throughput can grow as $m\log
\log n$ in the presence of either the primary broadcast or MAC
channel. Hence, the growth rate of the throughput is unaffected by the
presence of the primary (thus optimal). Furthermore, the interference
on the primary can also be made to decline to zero, while maintaining
the secondary average throughput to grow as $\Theta(\log \log n)$. 

\appendices

\section{Proof of Theorem~\ref{thm:AvgRMB}}
\label{Appendix:thmMB}

\begin{Proof}
We rewrite~\eqref{eq:InstantR} as
\begin{equation}
\RmacInstant = \log \det \bigg(I + \bfm{H}(\calm{S})
Q_s\bfm{H}^{\dag}(\calm{S}) \big(I + \bfm{G}_s
Q_p\bfm{G}_s^{\dag}\big)^{-1}\bigg) \label{eq:InstantRMB}
\end{equation}
Because for any positive definite matrix $A$ and $B$, the function
$\log \det (I + AB^{-1})$ is convex in $B$~\cite[Lemma
  \Rmnum{2}.3]{Diggavi2001}, we have
\begin{align}
\calm{R}_{mac} & =
\Ex_{\bfm{H}}\big[\Ex_{\bfm{G}_s}[\RmacInstant\,|\,
    \bfm{H}]\big] \label{eq:AvgRConditionalMB}
\\ & >      \Ex_{\bfm{H}}\bigg[\log \det \bigg(I + \bfm{H}(\calm{S})
Q_s\bfm{H}^{\dag}(\calm{S}) \big(I + \Ex[\bfm{G}_s
Q_p\bfm{G}_s^{\dag}]\big)^{-1}\bigg) \bigg] \label{eq:AvgRJensen1MB}
\\ & =      \Ex_{\bfm{H}}\bigg[\log \det \bigg(I + \frac{\rho_s}{1+P_p}\bfm{H}
(\calm{S})\bfm{H}^{\dag}(\calm{S}) \bigg)\bigg] \label{eq:AvgRJensen2MB}
\end{align}
where~\eqref{eq:AvgRJensen1MB} uses the Jensen inequality and the fact
that $\bfm{H}(\calm{S})$ and $\bfm{G}_s$ are independent. Substituting
$Q_p$ from~\eqref{eq:PUTxPowerB} and noting that
$\Ex[\bfm{G}_s\bfm{G}_s^{\dag}]=\Mprimary I_{m\times m}$, we
have~\eqref{eq:AvgRJensen2MB}.

Now we bound the right hand side of~\eqref{eq:AvgRJensen2MB}. Recall
that $|\calm{A}|$ and $|\calm{S}|$ are the random number of eligible
users and active users, respectively. By the Chebychev inequality, for
any $\epsilon>0$, we have
\begin{align}
\Pr\bigg( |\calm{A}| > (1-\epsilon) \bar{k}_s \bigg) & > 1-\frac{1-p}{\epsilon^2np}
\\ &= 1 - O\big(\bar{k}_s^{-1}\big) \label{eq:ChebyMB}
\end{align}
where in the above we use the fact $\bar{k}_s = np$. Then, we expand
 \eqref{eq:AvgRJensen2MB} based the event
$\{|\calm{A}|>(1-\epsilon)\bar{k}_s\}$ and its complement, and discard the
non-negative term associated with its complement:
\begin{align}
\calm{R}_{mac} & >  \Ex\bigg[\log \det \bigg(I +
  \frac{\rho_s}{1+P_p} \bfm{H}(\calm{S}) \bfm{H}^{\dag}(\calm{S})\bigg)\,\bigg | \,
  |\calm{A}| > (1-\epsilon)\bar{k}_s \bigg]\Pr\bigg(|\calm{A}| > (1-\epsilon)\bar{k}_s\bigg) \label{eq:AvgRConditionalLB1MB}
\\ & \ge  \Ex\bigg[\log \det \bigg(I +
  \frac{\rho_s}{1+P_p} \bfm{H}(\calm{S}) \bfm{H}^{\dag}(\calm{S})\bigg)\,\bigg | \,
  |\calm{A}|=(1-\epsilon)\bar{k}_s \bigg]
\bigg(1-O\big(\bar{k}_s^{-1}\big)\bigg)\label{eq:AvgRConditionalLB2MB}
\\ & =  \Ex\bigg[\log \det \bigg(I +
  \frac{\rho_s}{1+P_p} \bfm{H}(\calm{S}) \bfm{H}^{\dag}(\calm{S})\bigg)\,\bigg | \,
  |\calm{S}|=(1-\epsilon)\bar{k}_s \bigg] \bigg(1-O\big(\bar{k}_s^{-1}\big)\bigg)\label{eq:AvgRConditionalLB3MB}
\end{align}
where in the inequality~\eqref{eq:AvgRConditionalLB2MB}, we apply the
result in~\eqref{eq:ChebyMB} and the fact that the conditional
expectation of the right hand side of~\eqref{eq:AvgRConditionalLB1MB}
is non-decreasing in $|\calm{A}|$. Since $|\calm{S}| = (1-\epsilon)\bar{k}_s
$ in case of $|\calm{A}| = (1-\epsilon)\bar{k}_s $, then we obtain
\eqref{eq:AvgRConditionalLB3MB} due to the average throughput
depending on $|\calm{A}|$ via the size of $\calm{S}$.

Recall that each entry of $\bfm{H}(\calm{S})$ is
i.i.d. $\calm{CN}(0,1)$. Conditioned on $|\calm{S}|=(1-\epsilon)\bar{k}_s$,
$\bfm{H}(\calm{S}) \bfm{H}^{\dag}(\calm{S})$ is a Wishart Matrix with
degrees of freedom $(1-\epsilon)\bar{k}_s$, we have~\cite[Theorem
  1]{Bertrand2004}
\begin{align}
\calm{R}_{mac} & > \bigg(m \log \big(1+ \frac{(1-\epsilon)\rho_s\bar{k}_s}{1+P_p}\big)
+ O\big(\bar{k}_s^{-1}\big) \label{eq:AvgRHardenLB1MB}
\bigg)\bigg(1-O\big(\bar{k}_s^{-1}\big)\bigg)
\\ & = m \log \big(1+ \frac{(1-\epsilon)\rho_s\bar{k}_s}{1+P_p}\big) + O\big(\frac{\log
  \bar{k}_s}{\bar{k}_s}\big)  \label{eq:AvgRHardenLB2MB}
\\ & = m \log \rho_s\bar{k}_s + m\log (1-\epsilon) - m\log (1+P_p)+ O\big(\frac{\log
  \bar{k}_s}{\bar{k}_s}\big)  \label{eq:AvgRHardenLB3MB}
\end{align}
Since the above inequality holds for any $\epsilon>0$, we have
\begin{equation}
\calm{R}_{mac} \ge m \log \rho_s\bar{k}_s - m\log (1+P_p)+
O\big(\frac{\log \bar{k}_s}{\bar{k}_s}\big)  \label{eq:AvgRLBMB}
\end{equation}

Now we find an upper bound for $\calm{R}_{mac}$. For convenience, we
denote
\begin{equation}
R_{mac,0} = \log \det \bigg(I + \rho_s \bfm{H}(\calm{S})
\bfm{H}^{\dag}(\calm{S}) + \bfm{G}_s\, Q_p\,
\bfm{G}_s^{\dag}\bigg) \label{eq:InstantRmac0MB}
\end{equation}
and 
\begin{equation}
R_I = \log \det \bigg(I + \bfm{G}_s
\, Q_p\, \bfm{G}_s^{\dag}\bigg) \label{eq:InstantRIMB}
\end{equation}
So the average throughput can be written as
\begin{align}
\calm{R}_{mac} & = \Ex\big[R_{mac,0}\big] - \Ex\big[R_I\big] \label{eq:AvgRMB}
\end{align}

Using the inequality $\det(A)\le
\big(\trace(A)/k)\big)^k$~\cite{Coverbook}, where $A$ is a $k\times
k$ positive definite matrix, $R_{mac,0}$ is bounded by
\begin{equation}
R_{mac,0} \le m\log
  \bigg(1+\frac{1}{m}\trace \bigg(\rho_s \bfm{H}(\calm{S})
  \bfm{H}^{\dag}(\calm{S}) + \bfm{G}_s\, Q_p\,
  \bfm{G}_s^{\dag}\bigg)\bigg)
\end{equation}
Therefore,
\begin{align}
\Ex[R_{mac,0}] & \le m\Ex\bigg[\log
  \bigg(1+\frac{1}{m}\trace \bigg(\rho_s \bfm{H}(\calm{S})
  \bfm{H}^{\dag}(\calm{S}) + \bfm{G}_s\, Q_p\,
  \bfm{G}_s^{\dag}\bigg)\bigg)\bigg] 
\\& \le m \log\bigg(1+\frac{\rho_s}{m}\Ex\big[\trace\big(\bfm{H}(\calm{S})
  \bfm{H}^{\dag}(\calm{S})\big)\big] +
\frac{1}{m}\Ex\big[\trace\big(\bfm{G}_s
  Q_p\bfm{G}_s^{\dag}\big)\big]\bigg) \label{eq:AvgR0TraceUBMB}
\\ &\le m \log\big(1+\rho_s \bar{k}_s + P_p\big) \label{eq:AvgR0UBMB}
\end{align}
where~\eqref{eq:AvgR0TraceUBMB} uses the Jensen inequality. To obtain
the inequality~\eqref{eq:AvgR0UBMB}, we use the facts that
$\Ex\big[\trace\big(\bfm{G}_s Q_p\bfm{G}_s^{\dag}\big)\big]=P_p$ by
substituting $Q_p$ given by~\eqref{eq:PUTxPowerB} as well as
$\Ex\big[\trace\big(\bfm{H}(\calm{S})
  \bfm{H}^{\dag}(\calm{S})\big)\big]\le m\bar{k}_s$ due to $|\calm{S}|\le
\bar{k}_s$. 

Now we lower bound the second term
in~\eqref{eq:AvgRMB}. From~\cite[Theorem 1]{Oyman2003}, we have
\begin{align}
\Ex[R_I]& \ge m_{\sfm{min}}\log \bigg(1 +
\frac{P_p}{\Mprimary }\exp\bigg(\frac{1}{m_{\sfm{min}}}\sum_{j=1}^{m_{\sfm{min}}}\sum_{i=1}^{m_{\sfm{max}}-j}\frac{1}{i}-\gamma\bigg)\bigg) \label{eq:AvgRILBMB}
\\ &\stackrel{\Delta}{=} \calm{R}_I \label{eq:AvgR_IMB}
\end{align}
where $m_{\sfm{min}} = \min(m,\Mprimary )$, $m_{\sfm{max}} = \max(m,\Mprimary )$ and
$\gamma$ is the Euler's constant. Notice that $\calm{R}_I$ is a finite
constant independent of $n$ and $\Gamma$.

Combining~\eqref{eq:AvgR0UBMB} and~\eqref{eq:AvgR_IMB}, we have
\begin{align}
\calm{R}_{mac} & \le m\log (1+\rho_s \bar{k}_s+P_p) - \calm{R}_I  \label{eq:AvgRUBMB}
\end{align}

Finally, substituting $\bar{k}_s$ given by~\eqref{eq:k_sMB} and noting
that $\bar{k}_s=\Theta(n^{\frac{1}{\Nprimary +1}})$, we have
\begin{align}
 \calm{R}_{mac} & \ge \frac{m}{\Nprimary +1}\log n + \frac{1}{\Nprimary +1}\log
 \big(\rho_s\Gamma^{\Nprimary }\big)- m\log(1+P_p) +
 O\big(n^{-\frac{1}{\Nprimary +1}}\log n \big) 
\\ \calm{R}_{mac}  & \le  \frac{m}{\Nprimary +1}\log n + \frac{1}{\Nprimary +1}\log \big(\rho_s\Gamma^{\Nprimary }\big)- \calm{R}_I + O\big(n^{-\frac{1}{\Nprimary +1}}\big)
\end{align}
where we use the identity $\log (x+y) = \log x + \log(1+x/y)$ in the
above inequalities. This completes the proof.
\end{Proof}

\section{Proof of Theorem~\ref{thm:Optimality}}
\label{Appendix:thmOpt}

\begin{Proof}
We develop an upper bound for the secondary throughput in the presence
of the primary broadcast only; the development is similar in the
presence of the primary MAC and thus is omitted. We consider an
arbitrary active user set $\calm{S}$ and transmit covariance matrix
given by~\eqref{eq:SUTxSigM}, such that the interference constraints
on the primary are satisfied.

By removing the interference from the primary to the secondary, the
secondary throughput is enlarged. Then, using the inequality $\det(A)
\le \big(\trace(A)/k\big)^k$~\cite{Coverbook}, where $A_{k\times k}$
is a positive definite matrix, we have
\begin{align}
\RmacInstant & \le m \log
\bigg(1+\frac{1}{m}\trace\big(\bfm{H}(\calm{S})Q_s\bfm{H}^{\dag}(\calm{S})\big)\bigg) \label{eq:UPBoundWithoutInt}
\end{align}
Let $\bfm{h}_i$ be the $m\times 1$ vector of channel coefficients from
the secondary user $i$ ($i\in\calm{S}$) to the secondary base station,
corresponding to a certain column of $\bfm{H}(\calm{S})$. Since $Q_s$
is diagonal, we have
\begin{align}
\trace\big(\bfm{H}(\calm{S})Q_s\bfm{H}^{\dag}(\calm{S})\big) 
&= \sum_{i\in\calm{S}}  \rho_i \,\trace\big(\bfm{h}_i\bfm{h}_i^{\dag}\big) 
\\&=\sum_{i\in\calm{S}} \rho_i \,|\bfm{h}_i|^2 
\\& \le \max_{i\in \calm{S}} |\bfm{h}_i|^2 \, \sum_{i\in\calm{S}} \rho_i
\\ & \le\max_{1\le i\le n} |\bfm{h}_i|^2 \,\sum_{i\in\calm{S}} \rho_i
\end{align}
where $\rho_i$ is the transmit power of the secondary user $i$. Let
\begin{equation}
\PSecondary_{sum} = \sum_{i\in\calm{S}} \rho_i
\end{equation}
and 
\begin{equation}
\hmax = \max_{1\le i\le n} |\bfm{h}_i|^2 
\end{equation}
We can rewrite the right hand side of~\eqref{eq:UPBoundWithoutInt} as
\begin{equation}
\RmacInstant \le m\log
\big(1+\frac{1}{m}\hmax\PSecondary_{sum}\big) \label{eq:UPBoundMB}
\end{equation}

We first bound $\PSecondary_{sum}$ and formulate an optimization as:
\begin{align}
&\max_{\calm{S},\,\{\rho_i\}}\, \PSecondary_{sum} \nonumber
\\ s.t.: &\  \rho_i\le \rho_s \  \text{for}\ i\in \calm{S},  \nonumber
\\& \big[\bfm{G}_p\, Q_s\, \bfm{G}_p^{\dag}\big]_{\ell,\ell} \le
\Gamma \ \text{for}\ 1\le\ell\le \Nprimary \label{eq:Optimal1} 
\end{align}
which is a standard linear programming, and the solution is denoted by
$\PSecondary_{sum}^*$. Then, $\PSecondary_{sum}^*$ is the maximum total
transmit power, depending on the channel realizations for each
transmission.

Subject to the interference constraints on the primary, the user
selection and power allocation are coupled, and a direct analysis is
difficult. Instead, we will find an upper bound for
$\PSecondary_{sum}^*$. Notice that the total interference (on all
primary users) caused by the secondary user $i$ is $\rho_i
|\bfm{g}_{p,i}|^2$, where $\bfm{g}_{p,i}$ is the vector of channel
coefficients from the secondary $i$ to all $\Nprimary$ primary
users. We relax the set of individual interference constraints
in~\eqref{eq:Optimal1} with a single sum interference constraint:
\begin{align}
\sum_{i\in \calm{S}}\rho_i |\bfm{g}_{p,i}|^2 \le \Nprimary  \Gamma \label{eq:SumIntConstraint}
\end{align}
Notice that $\bfm{g}_{p,i}$ corresponds to a certain column in
$\bfm{G}_p$.

Order the cross channel gains $\{|\bfm{g}_{p,i}|^2\}_{i=1}^n$ of all
the secondary users and denote the ordered cross channel gains by
\begin{equation}
 |\tilde{\bfm{g}}_{p,1}|^2\le|\tilde{\bfm{g}}_{p,2}|^2\le\cdots\le|\tilde{\bfm{g}}_{p,n}|^2
\end{equation}
Then, we further relax the sum interference
constraint~\eqref{eq:SumIntConstraint} by replacing
$\{|\bfm{g}_{p,i}|^2\}_{i\in\calm{S}}$ with the first $|\calm{S}|$
smallest cross channel gains
$\{|\tilde{\bfm{g}}_{p,i}|^2\}_{i=1}^{|\calm{S}|}$. Thus, we have:
\begin{align}
&\max_{\calm{S},\,\{\rho_i\}}\, \PSecondary_{sum}\nonumber
\\ \text{s.t.:} &\sum_{i=1}^{|\calm{S}|} \rho_i |\tilde{\bfm{g}}_{p,i}|^2 \le
\Nprimary \Gamma  \nonumber 
\\ & \rho_i\le \rho_s   \  \text{for}\ 1\le i \le |\calm{S}| \label{eq:OptP_sum1}
\end{align}
For any channel realizations, the solution for the above problem,
denoted by $P_{sum,1}^*$, is always greater than, or equal to
$\PSecondary_{sum}^*$. Notice that $P_{sum,1}^*$ is also a random
variable. Since $\{|\tilde{\bfm{g}}_{p,i}|^2\}$ is in non-decreasing in
$i$, the set of $\{\rho_i\}$ that achieves $P_{sum,1}^*$ satisfies
$\rho_i\ge \rho_j$, for $i\le j$. In other words, we have $\rho_i =
\rho_s$, for $i=1$ to $|\calm{S}|-1$, and $\rho_{i} \le \rho_s$, for
$i=|\calm{S}|$. 

Let $\Smax$ be the maximum value of $|\calm{S}|$ that
satisfies the constraint
\begin{equation}
\rho_s\sum_{i=1}^{|\calm{S}|-1} |\tilde{\bfm{g}}_{p,i}|^2\le \Nprimary  \Gamma \label{eq:ConstraintOpt3}
\end{equation}
We have
\begin{equation}
P_{sum,1}^* \le \rho_s \Smax \label{eq:S_max}
\end{equation}
where in~\eqref{eq:S_max} we have an inequality, because the
constraint~\eqref{eq:ConstraintOpt3} is relaxed by discarding
$\rho_{|\calm{S}|}$ compared to the interference constraint
in~\eqref{eq:OptP_sum1} .

Now, we focus on bounding $\rho_s\Smax$. For any positive
integer $k$, we have
\begin{equation}
\Pr\big(\Smax < k\big) \ge \Pr\big(\sum_{i=1}^{k-1}
|\tilde{\bfm{g}}_{p,i}|^2 >\frac{\Nprimary \Gamma}{\rho_s} \big) \label{eq:S_maxRelation}
\end{equation}
which comes from the fact that the event of the right hand side
implies the event of the left hand side. Notice that $\sum_{i=1}^{k-1}
|\tilde{\bfm{g}}_{p,i}|^2$ is a sum of least order statistics out of
$\{|\bfm{g}_{p,i}|^2\}_{i=1}^n$ with i.i.d. Gamma$(\Nprimary ,1)$
distributions. We apply some results in the development
of~\cite[Proposition 12]{Mitran2010}, and obtain\footnote{For our
  case, $\frac{1}{\lambda}=\gamma = \Nprimary $.}
\begin{equation}
 \Pr\big(\sum_{i=1}^{f(n)-1} |\tilde{\bfm{g}}_{p,i}|^2 >\frac{\Nprimary \Gamma}{\rho_s}
 \big) > 1- O\big(\frac{1}{f(n)}\big)\label{eq:Mitran}
\end{equation}
where $f(n) = c_0\, n^{\frac{1}{\Nprimary +1}}$, and $c_0 = \big(\frac{\Gamma
  (\Nprimary +1)}{(1-\epsilon)\rho_s}\Nprimary ^{-\frac{1}{\Nprimary }}\big)^{\frac{\Nprimary }{\Nprimary +1}}$.
For large $\Nprimary $ and small $\epsilon$, $c_0\approx \frac{\Gamma}{\rho_s}(\Nprimary +1)$.

Let $k=f(n)$ in~\eqref{eq:S_maxRelation} and combine
with~\eqref{eq:Mitran}:
\begin{align}
\Pr\bigg(\rho_s\Smax<  \rho_s\, f(n)  \bigg)
> 1-O\big(n^{-\frac{1}{\Nprimary +1}}\big) \label{eq:S_maxBound}
\end{align}

After characterizing $\rho_s \Smax$, now we return to
$\PSecondary_{sum}^*$. To simplify notation, we denote
\begin{equation}
\bar{p}_{sum} = \rho_s\, f(n) 
\end{equation}
Because $\PSecondary_{sum}^*\le \PSecondary_{sum,1}^* \le \rho_s
\Smax$ for any channel realizations,
from~\eqref{eq:S_maxBound}, we have
\begin{align}
\Pr\bigg(\PSecondary_{sum}^*\ge
\bar{p}_{sum}\bigg)&=1-\Pr\bigg(\PSecondary_{sum}^*<  \bar{p}_{sum}
\bigg) \nonumber
\\& < 1- \Pr\bigg(\rho_s\Smax <  \bar{p}_{sum}
\bigg) \nonumber
\\&<O\big(n^{-\frac{1}{\Nprimary +1}}\big) \label{eq:P_sumProb}
\end{align}

Now, we complete the analysis of $\PSecondary_{sum}^*$, and move to
$\hmax$. Because $\{|\bfm{h}_i|^2\}_{i=1}^n$ have i.i.d.  Gamma$(m,1)$
distributions, using the similar arguments developed in
Lemma~\ref{lemma:SINRBoundsBB}, we obtain
\begin{align} 
&\Pr\bigg( \hmax> \zeta_n\bigg) =
O\big(\frac{1}{\log n}\big) \label{eq:H_nProb}
\\ &\Ex\big[\hmax\,\big|\, \hmax >\zeta_n\big]<O(n\log
n) \label{eq:H_nConditionalMean}
\end{align}
where $\zeta_n$ is a deterministic sequence satisfying
\begin{equation}
\zeta_n=\log n + m\log\log n + O(\log \log \log n)
\end{equation}
Now we are ready to develop the upper bound for the
secondary throughput. Since $\PSecondary_{sum}\le \PSecondary_{sum}^*$,
from~\eqref{eq:UPBoundMB}, we have
\begin{align}
\calm{R}_{mac} & \le m\Ex_{\bfm{H},\PSecondary}\bigg[\log\bigg(1+\frac{1}{m}\hmax\PSecondary_{sum}^*\bigg)\bigg] 
\\&\le m \Ex_{\bfm{H},\PSecondary}\bigg[\log
  \bigg(1+\frac{1}{m}\hmax \PSecondary_{sum}^*
  \bigg)\,\bigg|\,\PSecondary_{sum}^* <
  \bar{p}_{sum}\bigg]\Pr\big(\PSecondary_{sum}^* < \bar{p}_{sum}\big)
\nonumber 
\\ & \quad+ m \Ex_{\bfm{H},\PSecondary}\bigg[\log \bigg(1+\frac{1}{m}\hmax
  \PSecondary_{sum}^*
  \bigg)\,\bigg|\,\PSecondary_{sum}^*\ge
  \bar{p}_{sum}\bigg]\Pr\big(\PSecondary_{sum}^* \ge
\bar{p}_{sum}\big) \label{eq:UPBound_0}
\\ &\le m \Ex_{\bfm{H}}\bigg[\log \bigg(1+\frac{1}{m}\hmax \bar{p}_{sum}
  \bigg)\bigg]\cdot 1 \nonumber 
\\ &\quad + m \Ex_{\bfm{H}}\bigg[\log
  \bigg(1+\frac{1}{m}\hmax \rho_s n
  \bigg)\bigg]\cdot
O\big(n^{-\frac{1}{\Nprimary +1}}\big) \label{eq:UPBoundP_sum}
\\ &\le m \Ex_{\bfm{H}}\bigg[\log \bigg(1+\frac{1}{m}\hmax \bar{p}_{sum}
  \bigg)\,\bigg|\, \hmax\le \zeta_n\bigg]\Pr\big(\hmax\le
\zeta_n\big) \nonumber 
\\ &\quad + m \Ex_{\bfm{H}}\bigg[\log \bigg(1+\frac{1}{m}\hmax \bar{p}_{sum}
  \bigg)\,\bigg|\, \hmax> \zeta_n\bigg]\Pr\big(\hmax>
\zeta_n\big) \nonumber 
\\ &\quad + m \Ex_{\bfm{H}}\bigg[\log
  \bigg(1+\frac{1}{m}\hmax \rho_s n \bigg)\,\bigg|\,
  \hmax\le \zeta_n\bigg]\Pr\big(\hmax\le
\zeta_n\big)O\big(n^{-\frac{1}{\Nprimary +1}}\big) \nonumber
\\ &\quad + m \Ex_{\bfm{H}}\bigg[\log
  \bigg(1+\frac{1}{m}\hmax \rho_s n \bigg)\,\bigg|\,
  \hmax> \zeta_n\bigg]\Pr\big(\hmax>
\zeta_n\big)O\big(n^{-\frac{1}{\Nprimary +1}}\big)  \label{eq:UPBoundConditionalH_n}
\\ &\le m \log \bigg(1+\frac{1} {m}\, \zeta_n\,\bar{p}_{sum}
  \bigg)\cdot 1 \nonumber 
\\ &\quad + m \log \bigg(1+\frac{\bar{p}_{sum}}{m}\,
\Ex\big[\hmax\,\big|\, \hmax> \zeta_n\big]
\bigg)\Pr\big(\hmax>
\zeta_n\big)  \nonumber
\\ &\quad + m \log \bigg(1+\frac{1} {m} \zeta_n\, \rho_s n
\bigg) \cdot 1 \cdot O\big(n^{-\frac{1}{\Nprimary +1}}\big)  \nonumber
\\ &\quad + m \log \bigg(1+\frac{\rho_s n} {m}
\Ex\big[\hmax\,\big|\, \hmax> \zeta_n\big]
\bigg)\Pr\big(\hmax>
\zeta_n\big)O\big(n^{-\frac{1}{\Nprimary +1}}\big) \label{eq:UPBoundJensen}
\\ &\le m \log \bigg(1+\frac{1} {m}\,\zeta_n\, \bar{p}_{sum} \bigg) \nonumber
\\&\quad + m \log
\bigg(1+\frac{\bar{p}_{sum}} {m} O(n\log n) \bigg)O(\frac{1}{\log
  n})\nonumber
\\&\quad + m \log
\bigg(1+\frac{1}{m} \zeta_n\rho_s n \bigg)\, O\big(n^{-\frac{1}{\Nprimary +1}}\big) \nonumber
\\&\quad + m \log
\bigg(1+\frac{\rho_s n} {m} O(n\log n) \bigg)O(\frac{1}{\log
  n})O\big(n^{-\frac{1}{\Nprimary +1}}\big) \label{eq:UPBoundJensenSubstitute}
\end{align}
where the second term in~\eqref{eq:UPBoundP_sum} comes from
using~\eqref{eq:P_sumProb} as well as the fact that $\PSecondary_{sum}^*$
is upper bounded by $\rho_s n$. In~\eqref{eq:UPBoundJensen}, we apply
the Jensen inequality to obtain the second and fourth
terms. Using~\eqref{eq:H_nProb} and~\eqref{eq:H_nConditionalMean}, we
have the second and fourth terms
in~\eqref{eq:UPBoundJensenSubstitute}. Finally, by substituting
$\bar{p}_{sum}$ and $\zeta_n$, we obtain
\begin{equation}
 \calm{R}_{mac} \le \frac{m}{\Nprimary +1} \log n + O(\log\log n) \label{eq:UPBound}
\end{equation}
This concludes the proof of this theorem.
\end{Proof}

\section{Proof of Lemma~\ref{lemma:SINRBoundsBB}}
\label{Appendix:lemmaSINR}

\begin{Proof}
First, we prove~\eqref{eq:L1BB}. Let
$Z=|\bfm{h}^{\dag}_i\bfm{\phi}_j|^2$ and $Y=\theta \big(\sum_{k\ne
  j}|\bfm{h}^{\dag}_i \bfm{\phi}_j|^2 + |\bfm{g}_{s,i}|^2\big)$. Then,
$Z$ has the exponential distribution, and $Y$ has the
Gamma$\big((m+\Mprimary -1),\theta \big)$ distribution. We can write
\begin{equation}
\LowSINR_i = \frac{Z}{c+Y}
\end{equation}
where $c=\frac{m}{\rho}$. Conditioned on $Y$, the pdf of $\LowSINR_i$
is given by
\begin{align}
f_L(x)& =\int_0^{\infty}f_{L|Y}(x|y)f_Y(y)dy
\\ & = \int_0^{\infty} (c+y)e^{-(c+y)x}\times
\frac{y^{m+\Mprimary -1}e^{-y/\theta}}{(m+\Mprimary -1)!\,\theta^{m+\Mprimary }}dy
\\ & = \frac{e^{-cx}}{(1+\theta x)^{m+\Mprimary }} \big(c(1+\theta x)+\theta (m+\Mprimary -1)\big)
\end{align}
So the cdf of $\LowSINR_i$ is
\begin{align}
F_L(x) & =1- \int_x^{\infty}f_L(t)dt
\\ & = 1-\frac{e^{-cx}}{(1+\theta x)^{m+\Mprimary -1}} \label{eq:F_LBB}
\end{align}
We define a grow function as
\begin{align}
g_L(x) & = \frac{1-F_L(x)}{f_L(x)}
\\ & = \frac{1+\theta x}{c(1+\theta x) + \theta(m+\Mprimary -1)}
\end{align}
Since $\lim_{x\rightarrow \infty} g_L^{\prime}(x)=0$, the limiting
distribution of $\LowSINR_{max}=\max_{1\le i\le n} \LowSINR_i$
exists~\cite{Davidbook}:
\begin{equation}
\lim_{n\rightarrow\infty} \big(F_L(b_n+a_nx)\big)^n = e^{-e^{-x}}
\end{equation}
where $b_n=F_L^{-1}(1-1/n)$ and $a_n=g_L(b_n)$. In general, an exact
closed-form solution for $a_n$ and $b_n$ is intractable, but an
approximation can be obtained, which is sufficient for asymptotic
analysis. After manipulating~\eqref{eq:F_LBB}, we have
\begin{equation}
b_n = \frac{1}{c}\log n - \frac{m+\Mprimary -1}{c} \log\log n
+O\big(\log\log\log n\big) 
\end{equation}
 and thus
\begin{equation}
a_n = \frac{1}{c}+O\big(\frac{1}{\log n}\big)
\end{equation}
It is straightforward to verify
$\lim_{n\rightarrow\infty}\big(ng_L^{\prime}(b_n)\big)=\infty$, so we
apply the expansion developed in~\cite[Eq. (22)]{Uzgoren}
\begin{equation}
\big(F_L(b_n+a_nx)\big)^n=\exp\bigg(-\exp(-x+\Theta(\frac{x^2}{\log^2
  n})\big)\bigg) \label{eq:asyLExpandBB}
\end{equation}
 Let $x_1=-\log \log n$ and substitute $x_1$
 into~\eqref{eq:asyLExpandBB}, we obtain~\eqref{eq:L1BB}.

Now, we prove~\eqref{eq:U2BB} and~\eqref{eq:U2ConditionalBB}. Since
$\UpSINR_i$ is similar to $\LowSINR_i$, except that the denominator
now has the Gamma$\big(\Mprimary ,\theta \big)$
distribution. Following the same steps of
obtaining~\eqref{eq:asyLExpandBB}, we have the expansion of the cdf of
$\UpSINR_{max}$:
\begin{equation}
\big(F_U(d_n+c_nx)\big)^n=\exp\bigg(-\exp(-x+\Theta(\frac{x^2}{\log^2
  n})\big)\bigg) \label{eq:asyUExpandBB}
\end{equation}
where 
\begin{equation}
d_n =  \frac{1}{c}\log n - \frac{\Mprimary }{c} \log\log n
+O\big(\log\log\log n\big) 
\end{equation}
and 
\begin{equation}
c_n = \frac{1}{c} + O\big(\frac{1}{\log n}\big)
\end{equation}
 \eqref{eq:U2BB} follows by substituting $x_2=\log \log n$ into
 \eqref{eq:asyUExpandBB}.

Finally, because $\Ex[\UpSINR_{max}] < n\Ex[\UpSINR_i]$~\cite{Davidbook},
we have
\begin{align}
\Ex\bigg[ \UpSINR_{max} \,\bigg|\, \UpSINR_{max} > d_n+\frac{1}{c}\log\log n \bigg] &
\le \frac{ n\Ex[\UpSINR_i] }{\Pr\big( \UpSINR_{max} >
  d_n+\frac{1}{c}\log\log n \big)}
\\& = \Theta(n\log n)
\end{align}
where we use~\eqref{eq:U2BB} in the last equality.
\end{Proof}

\section{Proof of Theorem~\ref{thm:AvgRBB} }
\label{Appendix:thmBB}

\begin{Proof}
We first find a lower bound for the secondary average throughput
$\calm{R}_{bc}$. We condition on $\PSecondary=\rho$ and let $l_n = b_n
- \frac{\rho}{m}\log\log n $, where $b_n$ is given by
Lemma~\ref{lemma:SINRBoundsBB}. Using~\eqref{eq:AvgRBounds} and
Lemma~\ref{lemma:StochasticOrder}, the conditional throughput
$\calm{R}_{bc|\PSecondary}(\rho)$ can be bounded as
\begin{align}
\calm{R}_{bc|\PSecondary}(\rho)  & \ge m \Ex\bigg[\log\big(1+\LowSINR_{max}\big)\,\bigg|\,\PSecondary=\rho\bigg] \label{eq:RConditionalStochasticLBBB}
\\ & \ge m \Ex\bigg[\log\big(1+\LowSINR_{max}\big)\,\bigg|\, \LowSINR_{max} \ge
  l_n,\,\PSecondary=\rho\bigg]\Pr\big( \LowSINR_{max} \ge l_n\,\big|\,\PSecondary=\rho \big) \label{eq:RConditionalLBExpand}
\\ & > m \bigg(\log\big(\frac{\rho}{m}\log n \big)+O\big(\frac{\log\log
n}{\log n}\big)\bigg) \bigg(1-\Theta\big(n^{-1}\big)\bigg) \label{eq:RConditionalLBSubstituted}
\\ & = m \log \big(\frac{\rho}{m}\log n \big) + O\big(\frac{\log\log
n}{\log n}\big) \label{eq:RConditionalLB}
\end{align}
From~\eqref{eq:RConditionalStochasticLBBB}
to~\eqref{eq:RConditionalLBExpand}, we discard the non-negative term
associated with the event $\{\LowSINR_{max} < l_n\}$. Using
 \eqref{eq:L1BB} from Lemma~\ref{lemma:SINRBoundsBB} and the
identity $\log(x+y) = \log x + \log (1+y/x)$, we
have~\eqref{eq:RConditionalLBSubstituted}.

Now we take the expectation with respect to $\PSecondary$. From
 \eqref{eq:SUTxPowerBB}, we have
\begin{equation}
\PSecondary > \frac{m\Gamma}{\max_{1\le i \le \Nprimary } |\bfm{g}_{p,i}^{\dag}|^2 +
  m\Gamma/P_s} \label{eq:PVariationBB} 
\end{equation} 
where $\bfm{g}_{p,i}^{\dag}$ is the $1\times m$ vector of channel
coefficients from the secondary base station to the primary user
$i$. Let the pdf of $\max_{1\le i \le \Nprimary } |\bfm{g}_p(i)|^2$ be
$f_{g_p}(x)$.  Because the random variable $\PSecondary$ is
(stochastically) greater than the right hand side
of~\eqref{eq:PVariationBB}, from Lemma~\ref{lemma:StochasticOrder}
and~\eqref{eq:RConditionalLB}, we have
\begin{align}
\calm{R}_{bc} & > \int_0^{\infty} m \log \bigg(\frac{\Gamma \log n}{x +
  m\Gamma/P_s} \bigg)f_{g_p}(x)\;dx  + O\bigg(\frac{\log\log n}{\log n}\bigg) 
\\ & \ge m \log \bigg(\frac{\Gamma \log   n}{\tilde{\mu}_1 + m\Gamma/P_s} \bigg) 
+ O\bigg(\frac{\log\log n}{\log n}\bigg) \label{eq:AvgRJensenLBBB}
\\ & = m \log\big(\Gamma \log   n \big) -m\log\big(\tilde{\mu}_1 +
m\Gamma/P_s\big) + O\bigg(\frac{\log\log n}{\log n}\bigg) \label{eq:AvgRLBBB}
\end{align}
where~\eqref{eq:AvgRJensenLBBB} comes from the convexity of
$\log(a+\frac{b}{x+c})$ and
\begin{equation}
\tilde{\mu}_1 = \Ex[\max_{1\le i \le \Nprimary } |\bfm{g}_p(i)|^2 ]
\end{equation}

To find an upper bound, we still begin with the conditional throughput
$\calm{R}_{bc|\PSecondary}(\rho)$. Let $u_n=d_n + \frac{\rho}{m}\log\log
n$, where $d_n$ is given by Lemma~\ref{lemma:SINRBoundsBB}. Then
\begin{align}
\calm{R}_{bc|\PSecondary}(\rho) & \le m
\Ex\bigg[\log\big(1+\UpSINR_{max}\big)\,\bigg|\,\PSecondary=\rho\bigg] \label{eq:RConditionalStochasticUBBB}
\\ & \le  m \Ex\bigg[\log\big(1+\UpSINR_{max}\big)\,\bigg|\,\UpSINR_{max} < u_n
  ,\,\PSecondary=\rho\bigg]\Pr\big(\UpSINR_{max} < u_n \big|\PSecondary=\rho\big) \label{eq:RConditionalUBExpand}
\\ & \quad +  m \Ex\bigg[\log\big(1+\UpSINR_{max}\big)\,\bigg|\,\UpSINR_{max} \ge u_n  ,\,\PSecondary=\rho\bigg]\Pr\big(\UpSINR_{max} \ge u_n\big|\PSecondary=\rho \big)  
\\ & < m\log (1+u_n)\big(1-\Theta\big(\frac{1}{\log n}\big)\big)\nonumber
\\ & \quad +  m  \log\big(1+\Ex[\UpSINR_{max}\,|\,\UpSINR_{max} \ge u_n
 ,\,\PSecondary=\rho \big]\big)\Theta\big(\frac{1}{\log n}\big)  \label{eq:RConditionalUBSubstituted}
\\ & < m\log(1+\frac{\rho}{m}\log n) + O(1) \label{eq:RConditionalUB}
\end{align}
where~\eqref{eq:RConditionalStochasticUBBB} comes
from~\eqref{eq:AvgRBounds}. We apply~\eqref{eq:U2BB} in
Lemma~\ref{lemma:SINRBoundsBB} and the Jensen inequality to
obtain~\eqref{eq:RConditionalUBSubstituted}. Using~\eqref{eq:U2ConditionalBB}
in Lemma~\ref{lemma:SINRBoundsBB} and substituting $u_n$, we
obtain~\eqref{eq:RConditionalUB}.

After calculating an upper bound for the conditional throughput, we
average over $\PSecondary$. From~\eqref{eq:SUTxPowerBB}, we have
\begin{equation}
\PSecondary \le \frac{m\Gamma}{\max_{1\le i \le \Nprimary } |\bfm{g}_{p,i}^{\dag}|^2}
\end{equation}
We denote
\begin{equation}
\frac{1}{\tilde{\mu}_2} = \Ex\big[ 1/\max_{1\le i \le \Nprimary } |\bfm{g}_{p,i}^{\dag}|^2 \big]
\end{equation}
Then, by the Jensen inequality, we have
\begin{align}
\calm{R}_{bc} & <  m\log\big(1+\frac{\log n}{m} \Ex[\PSecondary]\big) + O(1)
\\& <  m\log\big(1+\frac{\Gamma}{\tilde{\mu}_2}\log n\big) + O(1) \label{eq:UPBoundStochasticBB}
 \\ & = m\log(\Gamma\log n) - m\log \tilde{\mu}_2 + O(1)
\end{align}
where~\eqref{eq:UPBoundStochasticBB} holds since $\Ex[\PSecondary] \le
\frac{m\Gamma}{\tilde{\mu}_2}$. The theorem follows.
\end{Proof}

\bibliographystyle{IEEEtran} \bibliography{IEEEabrv,LiYang}

\end{document}